\documentclass[12pt]{report} 

\usepackage{amssymb}
\usepackage{graphicx}
\usepackage{multicol}
\usepackage{amsmath}
\usepackage{braket}
\usepackage{subcaption}
\usepackage{afterpage}
\usepackage{hyperref}
\hypersetup{
  colorlinks=true,
  citecolor=red,
  linkcolor=blue,
}

\def\mytitle{Loop ensembles in Stochastic Series Expansion of\\ Two-Dimensional Heisenberg Antiferromagnets}
\def\myname{Vedant R. Motamarri}
\def\mydegree{Dual Degree \\ (B.Tech. \& M.Tech.) }
\def\mysupervisor{Prof. Sumiran Pujari}
\def\myrollno{170260005}
\def\mydep{Department of Physics}

\begin{document}
 \baselineskip=18pt plus1pt
 \setcounter{secnumdepth}{3}
 \setcounter{tocdepth}{3}
 \pagenumbering{roman}


 \thispagestyle{empty}
\begin{center}
    { \Large {\bfseries {\mytitle}} \par}
\vspace{3\baselineskip}
    {\textit{submitted in partial fulfillment of the}\\
    \textit{requirements for the degree}}\par
\vspace{\baselineskip}
    {\textit{of} \par}
\vspace{\baselineskip}
    {\large \bf \mydegree \par} 
\vspace{\baselineskip}
    {\textit{by} \par}
\vspace{\baselineskip}
    {{\large {\bf \myname \\ \myrollno}} \par}
\vspace{1.5\baselineskip}
    {\it under the guidance of \par}
\vspace{\baselineskip}
    {{\large \bf \mysupervisor} \par}
\vspace{1.5\baselineskip}
    {\begin{figure}[!h] 
	\centering
	\includegraphics[width=35mm]{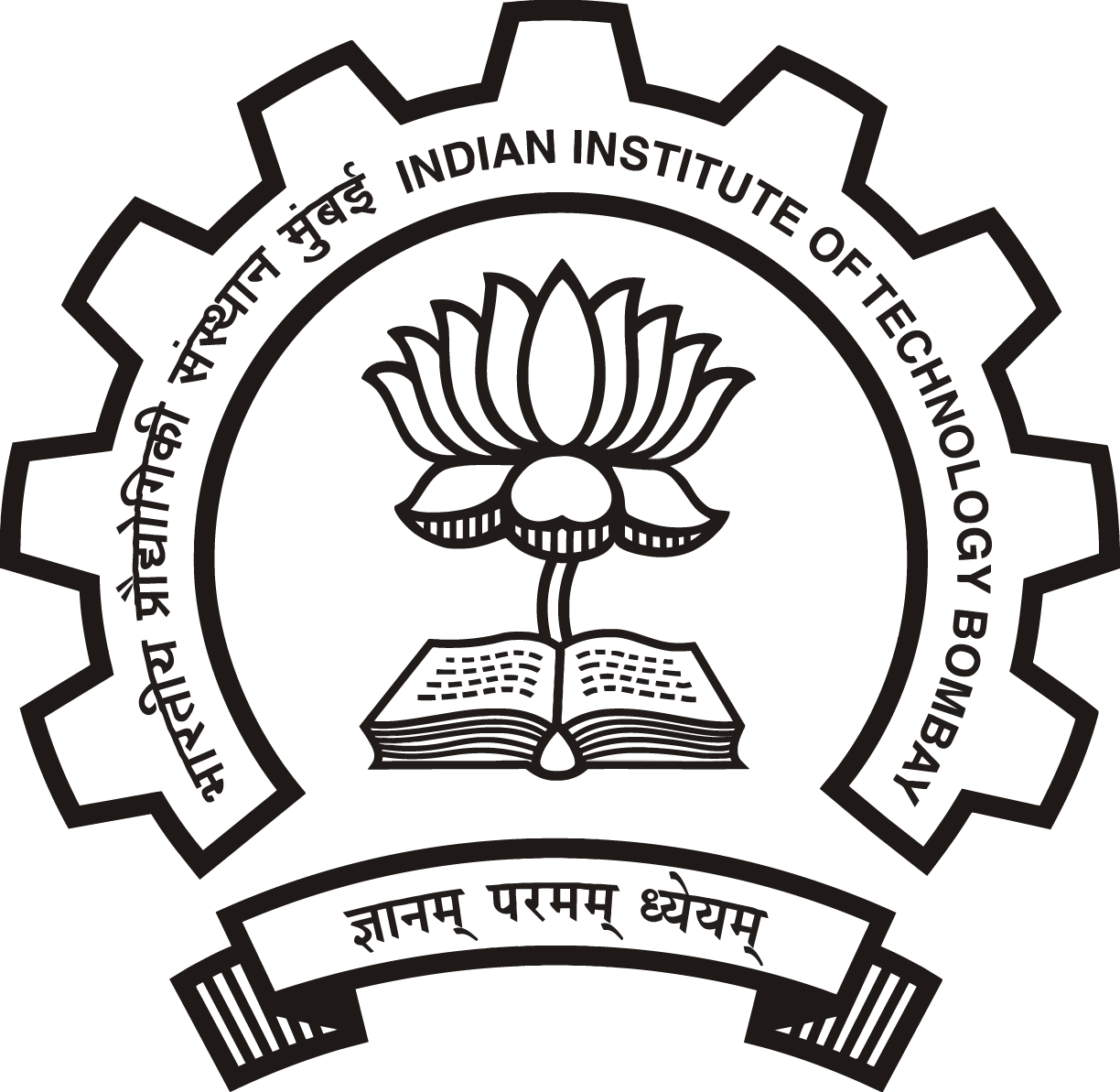} 
     \end{figure}
    }
\vspace{1.5\baselineskip}
    {\bf \MakeUppercase{\mydep} \par}
\vspace*{1ex}
    {\bf \uppercase{Indian Institute of Technology Bombay} \par}
 \end{center}
 \afterpage{\null \newpage}
\thispagestyle{plain}
\begin{center}
    \Large \textbf{\uppercase{Abstract}}
\end{center}

\vspace{3\baselineskip}


\noindent The Stochastic Series Expansion (SSE) method along with resummation over the spin or flavor
values maps the partition function of a quantum antiferromagnet to a closely-packed loop gas model
in one higher dimension. Earlier work by Nahum \textit{et al}.  has shown that certain closely-packed three-dimensional loop gas models exhibit phases dominated by macroscopic loops, wherein the corresponding joint distribution of loop lengths is Poisson-Dirichlet. On grounds of universality, the
same is expected of the ensemble of loops obtained in (2+1)-dimensional quantum antiferromagnets, albeit the loops emerge from a different microscopic origin. We sample the SSE loop ensemble for SU(N) antiferromagnets on a square lattice using Monte Carlo and study how the joint distribution varies with the degree of representation N and inverse temperature $\beta$. 
We observe that, for low temperatures and small $N (\leq 4)$, the distribution indeed shows characteristics of Poisson-Dirichlet behaviour when antiferromagnetic correlations dominate the system.
\vspace{\baselineskip}

Stochastic Series Expansion 
 \afterpage{\null \newpage}
 \tableofcontents
 \listoffigures
 
 \clearpage
 \pagenumbering{arabic}
 
 \chapter{Introduction \& Motivation}

Loops and random curves are objects of great interest in statistical physics and mathematics. Loop models have proven helpful in the study of critical phenomena in a wide range of problems including spin-models, polymers and Anderson-localized systems \cite{thesis}.
An interesting avenue of research where it's not apparent that loop models can play a role is the computational study of spin systems. Here loops can emerge when many-body systems are viewed as propagating in imaginary time either by path-integral or stochastic series expansion methods \cite{sandvik1,sandvik2}. In both these methods, the imaginary time axis adds an extra dimension to the problem and a $d$-dimensional quantum system can be related to a $(d+1)$-dimensional loop model. 

Previous work in mathematical and theoretical physics has strived to characterize the distribution of loop-lengths in such models \cite{ueltjmp,prl1}. Nearly two decades ago, D. Ueltschi and collaborators made a conjecture \cite{ueltarxiv} that the Poisson-Dirichlet distribution describes the properties of loops in three-dimensional systems. In the previous decade, in a series of papers A. Nahum \textit{et al}.\cite{prl1} gave a field theoretic derivation for this result and verified its applicability in the extended phase when a fraction of the lattice is covered in long-loops. 

In this project we study the properties of 2D quantum antiferromagnets in terms of 3d loop models, in light of the $d-(d+1)$ quantum-classical correspondence. Interestingly enough, in $\text{SU(N)}$ antiferromagnets, as one tunes the value of $N$, the system transitions from an extended/long-loop phase to a disordered short loop phase. One of the motivations for our work is to check whether the small $N$ extended phase is correctly described by the Poisson-Dirichlet distribution, otherwise derived for classical loop models. Another direction for enquiry is how the system makes the transition to the short loop phase and what distribution characterizes the properties of the loops when the Poisson-Dirichlet description is no longer applicable.

\section{Resummation based loops in Stochastic Series Expansion}

In the Stochastic Series Expansion technique, we are concerned with a Hamiltonian on a $d$-dimensional bipartite lattice
\begin{equation}
    H= J \sum_{\langle i,j \rangle} \mathbf{s}_i.\mathbf{s}_j
\end{equation}
For the SU(2) operator $\mathbf{s}$, one can tweak the Hamiltonian with a unitary rotation and rewrite it upto a constant as,
\begin{equation}
    H = -J \sum_{\langle i,j \rangle} H_{ij} \quad H_{ij} = \frac{1}{N} \sum_{\alpha \beta} \ket{\alpha_i \alpha_j} \bra{\beta_i \beta_j}, \label{H_sse}
\end{equation}
where $\ket{\alpha_i}$ are the spin-$z$ basis states at the site $i$. Note that qualitatively, this Hamiltonian acts on \textit{aligned} states $\ket{\alpha_i \alpha_j}$ and gives back other aligned states $\ket{\beta_i \beta_j}$, always with the same amplitude $1/N$. For SU(2) operators $N=2$, but Eq.(\ref{H_sse}) is written for general $N$ to emphasize the ease with which this formalism generalizes to higher representations. 

At a finite temperature, the partition function $Z=\text{Tr} \left \{\exp(-\beta H) \right \}$ can be expanded in Taylor series as $\sum_n \frac{(-\beta)^n}{n!} \text{Tr }H^n$. Writing out the trace operator more elaborately we get,
\begin{equation}
    Z = \sum_{n=0}^{\infty} \frac{(-\beta)^n}{n!} \sum_{S_n} \sum_{\alpha} \bra{\alpha} H_{b_1,\mu_1} H_{b_2,\mu_2} 
    ... H_{b_n,\mu_n} \ket{\alpha}.
\end{equation}
The form of the equation above suggests that the partition function is obtained by evaluating the matrix-element $\bra{\alpha} ... \ket{\alpha}$ over all possible combinations of $(S_n, \alpha)$. Here $\ket{\alpha}$ is a basis state of the lattice and $S_n$ is string of bond-operators that make up the Hamiltonian. $H_{b_l,\mu_l}$ is a an operator of \textit{kind} $\mu_l$ acting on bond $b_l$. The subscript $\mu_l$ specifies the particular operation performed i.e. the aligned two-site state ${\ket{\alpha}}$ which is taken to the state ${\ket{\beta}}$, as per Eq.(\ref{H_sse}).

\begin{figure}[h!]
    \centering
    \includegraphics[width=0.65\columnwidth]{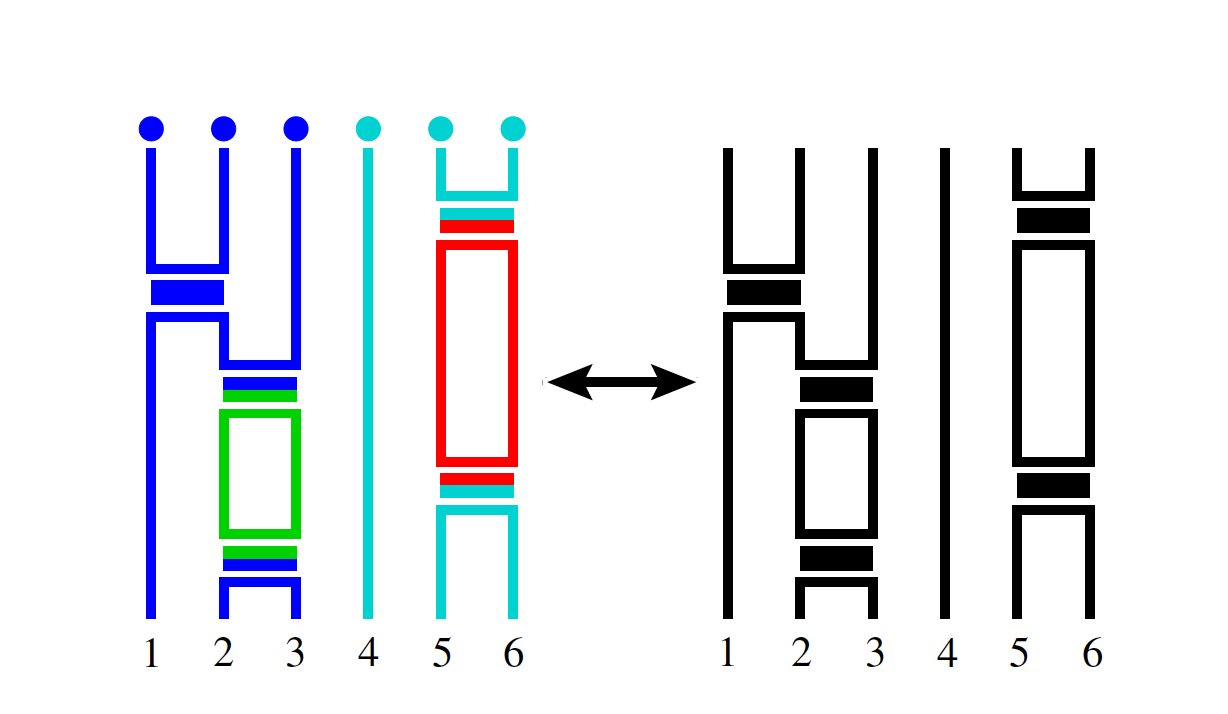}
    \caption{Operator string configurations for standard
SSE (\textit{left}) with colors denoting distinct spin orientations,  (\textit{right}) the ensemble characterized purely by uncolored loops. Ref.\cite{resumm}}
    \label{resumm}
\end{figure}

Pictorially, the pairing $(S_n, \alpha)$ translates to configurations of the type shown on the left in Fig.(\ref{resumm}). Each color indicates a different spin orientation amongst the $N$ possibilities, and $n$ bond operators make up the operator-string $S_n$ which manages to change the \textit{color} of the bond. The partition function is evaluated by summing over \textit{weights} of all such allowed configurations. The weight a $n$-step configuration is simply the contribution to $Z$ given by,
\begin{equation}
    \frac{(-\beta)^n}{n!} \bra{\alpha} H_{b_1,\mu_1} H_{b_2,\mu_2} 
    ... H_{b_n,\mu_n} \ket{\alpha}.
\end{equation}
which in our choice of the Hamiltonian Eq.(\ref{H_sse}) comes out as
\begin{equation}
    \frac{1}{n!} \left (\frac{\beta J}{N} \right )^n
\end{equation}
It is evident from Fig.(\ref{resumm}) that a different choice of coloring of the loops generates configurations with the same weights. One can thus consider instead a new ensemble with colorless loop-coverings shown on the right. This new description states that the partition function can be written as,
\begin{equation}
    Z = \sum_{C} W_C, \quad \text{with } W_C = \frac{N^{|C|}}{n!} \left (\frac{\beta J}{N} \right )^n.
\end{equation}
where the sum is over all possible loop-coverings $C$, each with a number of loops $|C|$.

The above described ensemble of loops on a $(d+1)$-dimensional lattice forms the backdrop of our work. We wish to explore the loop-length charecteristics in this language for different choices of $N$ and $\beta$ for $d=2$. For further details about Stochastic Series Expansion or about the resummation over loop colours, we guide the reader to the references \cite{resumm,sandvik1,sandvik2}.


 \chapter{Review of Literature -- Length distributions in 3D loop models}

The purpose of this chapter is to give an acquaintance with the key concepts that were used to derive the length distributions and associated quantities in 3D loop models. The ideas here follow from Ref.\cite{thesis}

\section{Loop models \& lattice field theory }
\begin{figure}
    \centering
    \includegraphics[width=\columnwidth]{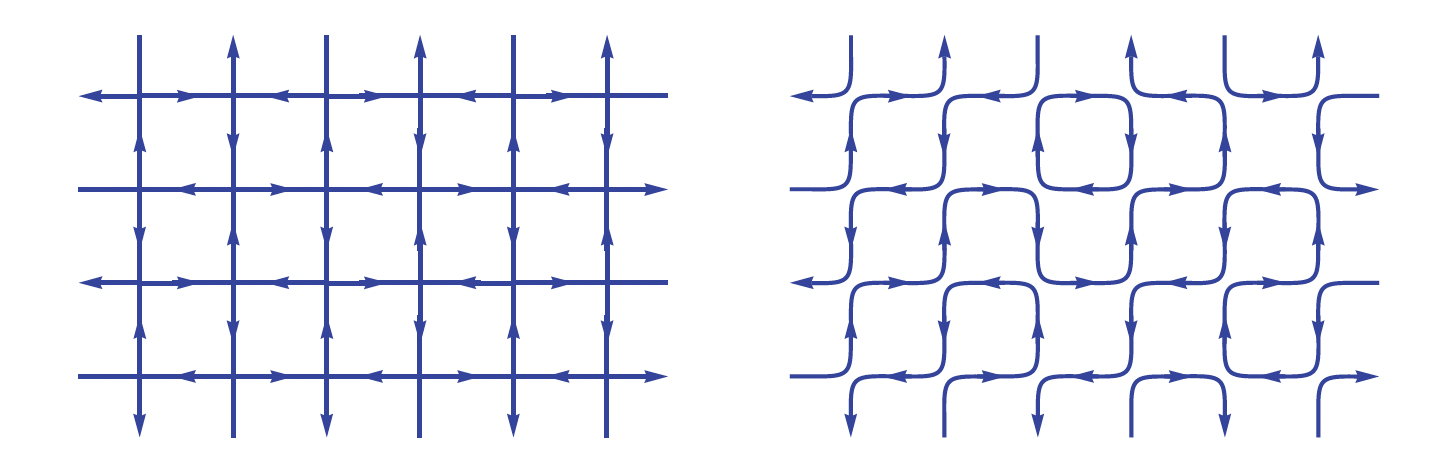}
    \caption{(\textit{left}) The 2D L lattice and (\textit{right}) a loop configuration obtained from an underlying pairing of links. Ref.\cite{thesis} }
    \label{2DL}
\end{figure}

A good starting point to consider is the four-coordinated directed lattice in 2D. Each link on such a lattice has an associated direction and configurations of loops are generated by picking a pairing at each node for the incoming and outgoing links. Assigning a probability $p$ to the left-turning pairing of links at a node, and $1-p$ otherwise, each configuration of loops has an associated weight $W_C$. See Fig.(\ref{2DL}) for illustration of one possible loop configuration in the 2D L lattice. 
If each loop has an additional degree of freedom $n$, called the \textit{fugacity}, which can be thought of as the number of possible colors allowed for the loops, the paritition function becomes:
\begin{equation}
    Z_{\text{loops}} = \sum_C n^{|C|} W_C , \qquad \text{with} \quad  W_C = p^{N_p}(1-p)^{N_{1-p}},
\end{equation}
where $|C|$ is the number of loops in a configuration, $N_p$ and $N_{1-p}$ indicate number of nodes with left and right-turning pairing respectively. The formulation described above can similarly be extended to 3D lattices. 

The applicability of field-theoretic methods to study the above loop model can be understood by adding new degrees of freedom $\mathbf{z}_l$ on each link of the lattice. The $\mathbf{z}_l$ are n-component complex vectors that satisfy
\begin{equation}
       \mathbf{z} = (z^1_l,z^2_l,..,z^n_l), \qquad \mathbf{z}^{\dagger} \mathbf{z}=n,
\end{equation}
and play the role of spins or lattice magnets. Using these new \textit{dof}s one can define an action
\begin{equation}
    \exp (-S_{\text{node}}) = p (\mathbf{z}_o^{\dagger} \mathbf{z}_i ) (\mathbf{z}_o'^{\dagger} \mathbf{z}_i' ) + 
    (1-p) (\mathbf{z}_o^{\dagger} \mathbf{z}_i' )(\mathbf{z}_o'^{\dagger} \mathbf{z}_i),
    \label{Snode}
\end{equation}
where both possible pairings of the outgoing links, denoted by subscripts ($o$,$o'$), to the incoming links ($i$,$i'$) are assigned different probabilities. The partition function for this loop model can be computed as
\begin{equation}
    Z = \text{Tr} \prod_{\text{nodes}} \exp(-S_{\text{node}}) = \sum_C W_C \prod _{\text{loops in C}} \text{Tr } (\mathbf{z}_1^{\dagger} \mathbf{z}_2 )(\mathbf{z}_2^{\dagger} \mathbf{z}_3 )...(\mathbf{z}_l^{\dagger} \mathbf{z}_i),
    \label{Sfull}
\end{equation}
which when simplified using $\text{Tr } 1 = 1$ and $\text{Tr } z^{\alpha}_l \bar{z}^{\beta}_l = \delta^{\alpha \beta}$, yields
\begin{equation}
    Z = \sum_C W_C \left (\prod_{\text{loops in C}} n \right)  = \sum_C n^{|C|} W_C = Z_{\text{loops}}
\end{equation}
Thus we see that there exists a correspondence between classical loop models and models with local magnetic degrees of freedom. The fugacity $n$ in the former model translates to number of components of the complex vector $\mathbf{z}$ in the latter model.

\begin{figure}[ht!]
    \centering
    \includegraphics[width=0.6\columnwidth]{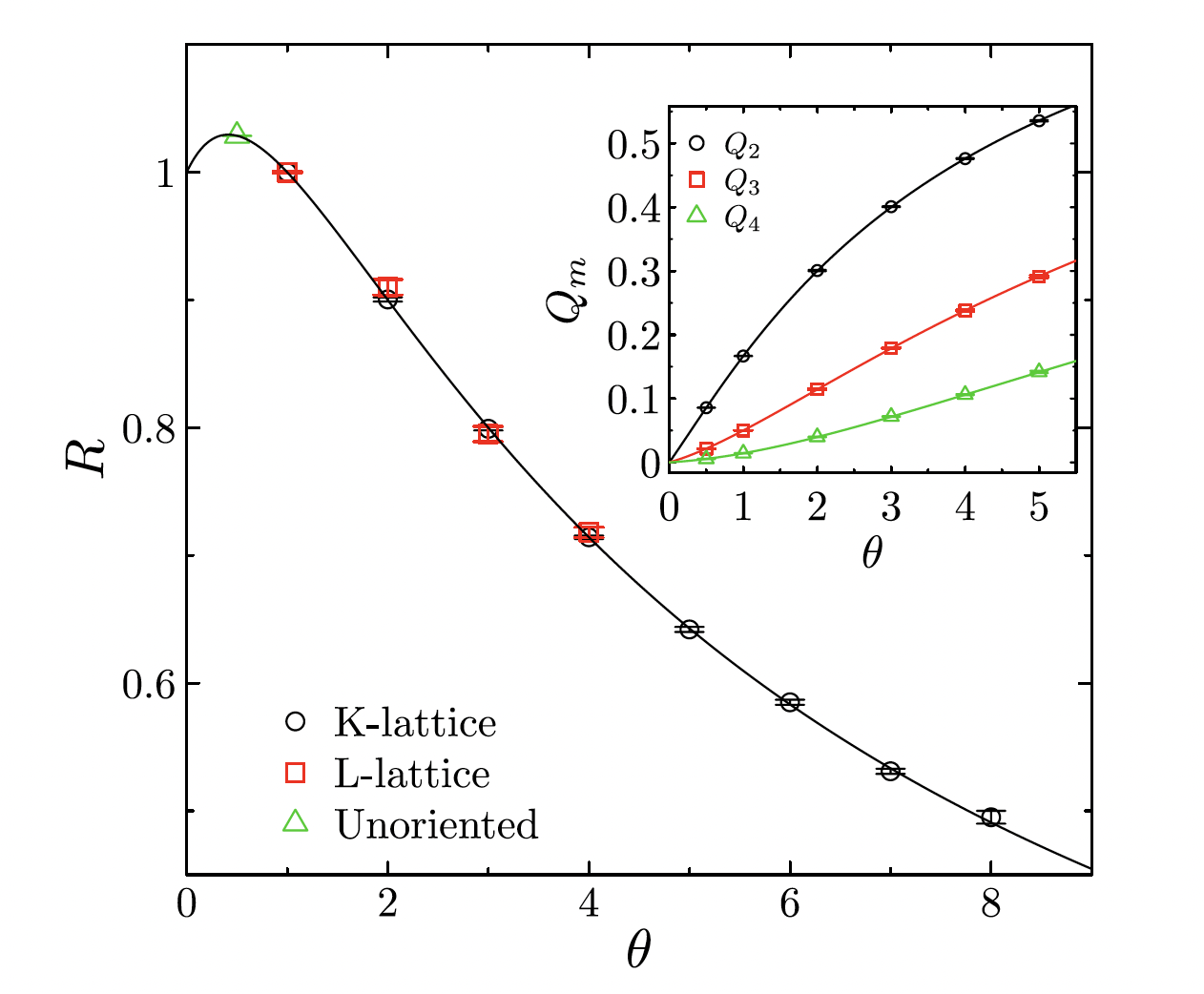}
    \caption{Comparison of simulation data (points) with theoretical values (lines) for ratios of moments of loop lengths. The main panel plots the one-loop ratio $R$ and the inset shows two-loop ratios $Q_m$. Ref. \cite{prl1}}
    \label{fig:my_label}
\end{figure}

\newpage

\section{Continuum description} 

To devise a continuum description of the above lattice field theory (in terms of the dof $\mathbf{z}_l$), one begins by identifying the symmetries of the model. The sum over loop configurations in Eq.(\ref{Sfull}) gives rise to a local U(1) gauge symmetry,
\begin{equation}
    \mathbf{z}_l \sim e^{i\phi} \mathbf{z}_l ,
\end{equation}
and the $\mathbf{z}^{\dagger} \mathbf{z}$ structure of the model implies a global SU(n) symmetry. The redundant overall phase $e^{i \phi}$ in the vector $\mathbf{z}$ implies that the spins live on the complex projective space $CP^{n-1}$, and thus a more suitable parametrisation is in terms of the gauge-invariant traceless Hermitian matrix
\begin{equation}
    Q^{\alpha \beta} = z^{\alpha} \bar{z}^{\beta} - \delta^{\alpha \beta}.
\end{equation}
Using the field $Q$, the authors devise for the continuum description, the $\sigma$ -model
\begin{equation}
    \mathcal{L} = \frac{1}{2g} \text{tr}(\nabla Q)^2
\end{equation}
where $Q$, in addition to being traceless \& Hermitian, also satisfies the constraint
\begin{equation}
    (Q+1)^2 = n (Q+1).
\end{equation}

\section{Moments of the distribution}

\begin{figure}[]
\begin{subfigure}{.51\columnwidth}
  \centering 
  \includegraphics[width=\linewidth]{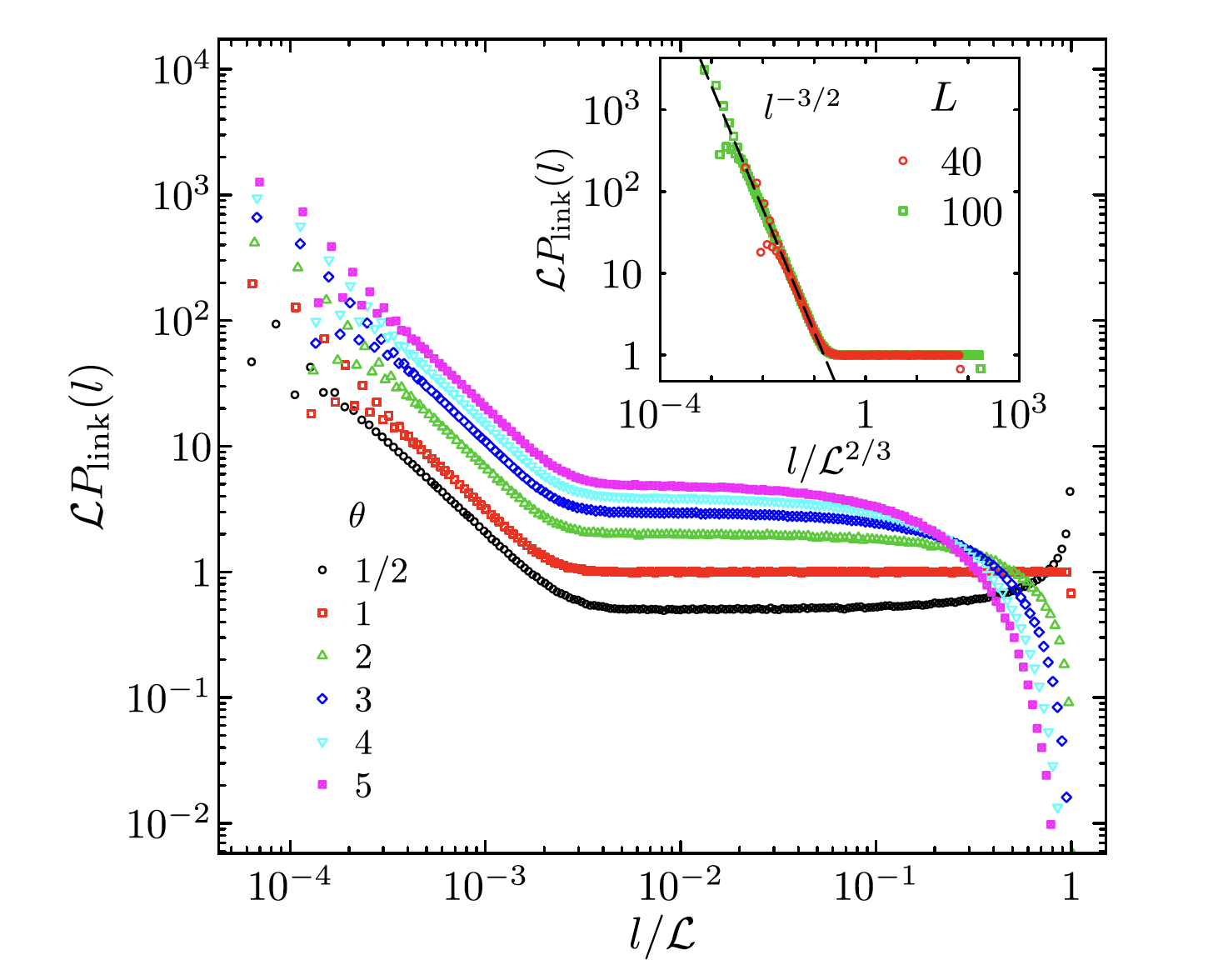} \caption{Distribution over all length scales}
  \label{fig:sub-first}
\end{subfigure}
\begin{subfigure}{.51\columnwidth}
  \centering
  \includegraphics[width=\linewidth]{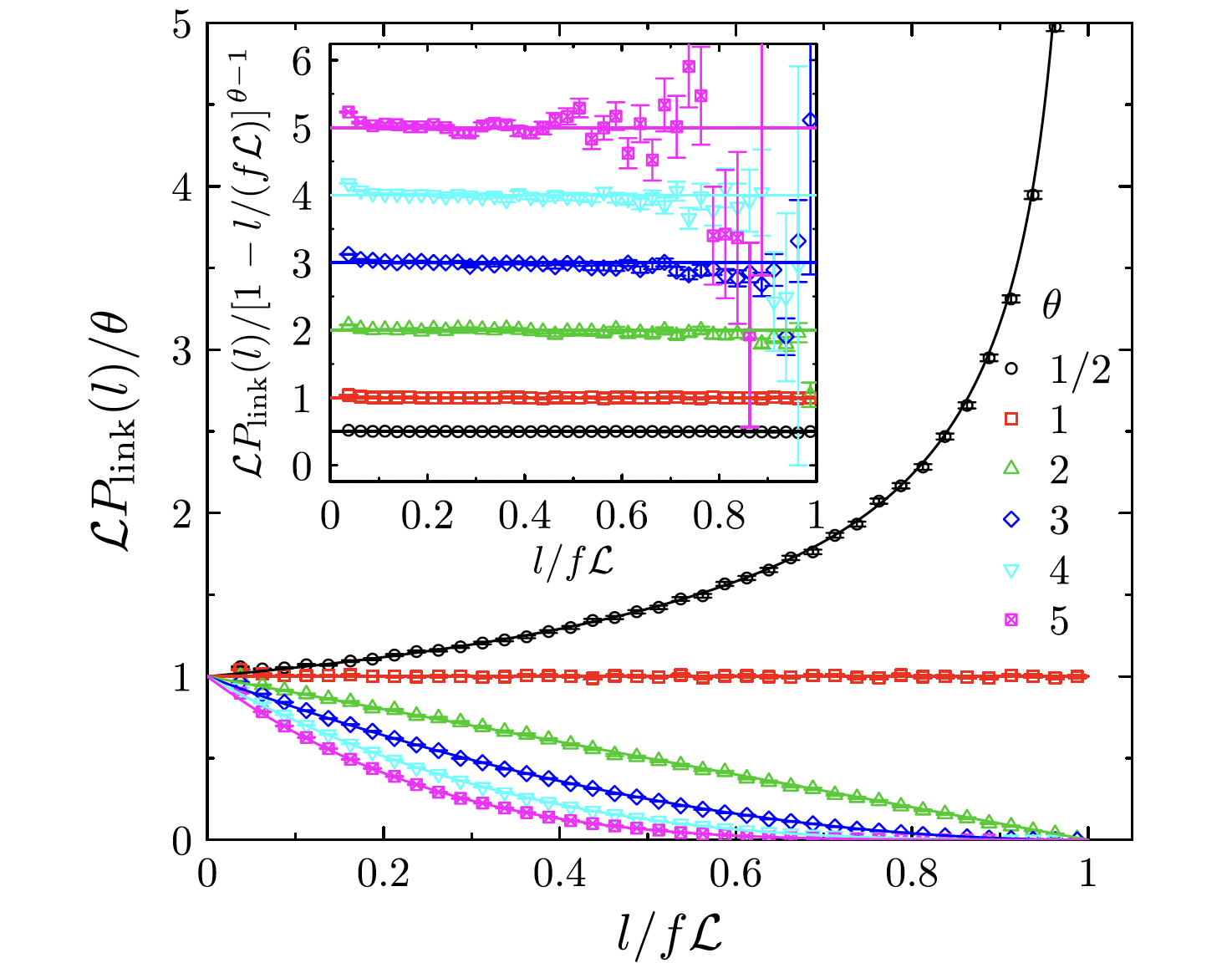}  
  \caption{Distribution in the PD regime}
  \label{fig:sub-second}
\end{subfigure}
\caption{Loop length distribution $P_{\text{link}}(l)$ for different values of fugacity $\theta$. Ref. \cite{prl1}}
\label{plink}
\end{figure}

The geometrical observables in the loop models, say for e.g. the probability for two loops to pass through the same point, are given in terms of correlation functions of Q. Using similar ideas, one can show that the moments of the length distribution are given as, 
\begin{equation}
    \mathcal{L}^{-m_{\text{tot}}} \left\langle \sum_{i_1,..,i_q}^{'}  l^{m_1}_{i_1}... l^{m_q}_{i_q} \right\rangle = f^{m_{\text{tot}}} 
    \frac{n^q \Gamma(n) \Gamma (m_1) .. \Gamma(m_q)}{\Gamma(n+m_{\text{tot}})}
\end{equation}
where $f$ is the fraction of the lattice covered with long loops and the total sum over all loop-lengths is $\mathcal{L}$. From this one computes ratios of moments of the distribution defined as,
\begin{equation}
        R = \frac{\langle \sum_i l^4_i \rangle}{ \langle \sum_k l^2_k \rangle^2} \quad \text{and} \quad 
    \mathcal{Q}_m = \frac{\langle \sum_{i\neq j} l_i^m l^m_j \rangle}{\langle \sum_k l^m_k \rangle^2}.
\end{equation}
which come out to be 
\begin{equation}
    R = \frac{6(\theta +1)}{(\theta+2)(\theta+3)} \quad \text{and } \quad
    \mathcal{Q}_m = \frac{(m-1+\theta)(m-2+\theta)..(\theta)}{(2m-1+\theta)(2m-2+\theta)...(m+\theta)}.
\end{equation}
where $\theta=n$ denotes the fugacity. 

Another key observable is the probability distribution $p_{\text{link}}(l)$ for the length $l$ of a loop passing through a randomly selected link. Considering a correlation length $\zeta$ which sets the scale beyond which loops are Brownian, for $\zeta \ll  l \ll L^2$, the loops are insensitive to the sample boundaries and display Brownian character
\begin{equation}
    p_{\text{link}}(l) = Cl^{-d/2}.
    \label{plink_brownian}
\end{equation}
For the longer loops in the system i.e. those with $L^2 \ll l \ll f \mathcal{L}$, the distribution carries over to the Poisson-Dirichlet form
\begin{equation}
    p_{\text{link}}(l) = \theta \mathcal{L}^{-1} [1 - l/(f \mathcal{L})]^{\theta -1}
    \label{plink_pd_eqn}
\end{equation}


 \chapter{Results \& Discussion}

\section{Tuning $N$}
\label{Ntx}

As the degree of the representation $N$ is increased, the system transitions from an extended phase, where long-loops proliferate, to a disordered phase with absence of macroscopic loops. We study below how the length distribution of loops passing through a randomly selected link $P_{\text{link}}(l)$, and the ratio of moments of the joint length distribution $R,Q_m$ vary over different values of $N$. The simulations are carried out for a 2D square lattice of side length $L=16$ with an inverse temperature $\beta=L$ to ensure existence of an extended phase for small $N$. The distribution is sampled using $10^5$ Monte Carlo steps.

\subsection{Loop length distribution}

We consider integer steps from $N=2$ to $9$. The key aspects of the distribution for consideration are as follows:

\begin{figure}[h!]
    \centering
    \includegraphics[width=1\columnwidth]{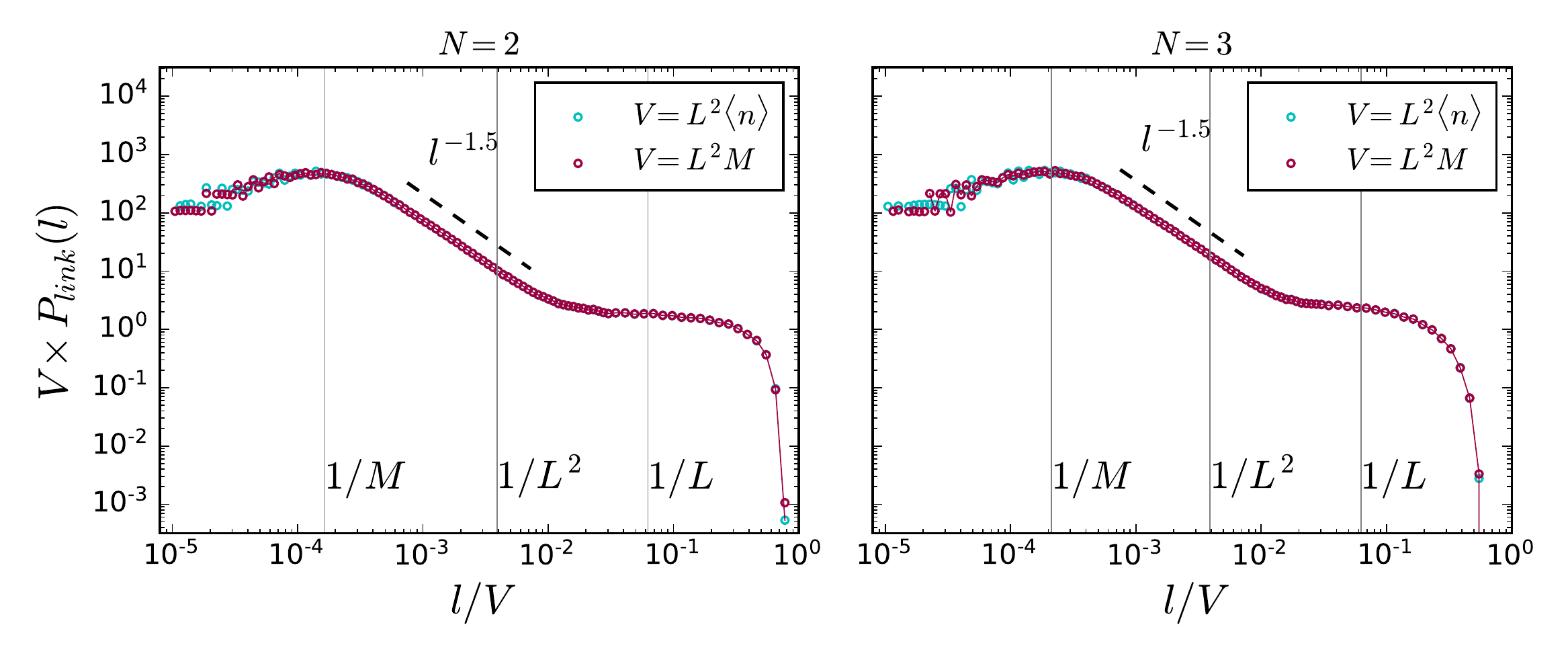}
    \caption{$V P_{\text{link}}(l)$ vs $l/V$ for $N=2$ and $N=3$, computed for SSE ensembles including (\textit{cyan}) and excluding (\textit{red}) identities}
    \label{N=23}
\end{figure}

\subsubsection*{Existence of long loops}As the value of $N$ increases one can track the maximum length of a loop in the ensemble, indicated by the value of maximum $l_{\text{max}}$ that carries a non-zero weight in the distribution. For $N=2$, where we expect an extended loop phase to exist, the ratio $l_{\text{max}}/V$ is close to 0.7. As $N$ is increased, the ratio keeps decreasing and there are no loops larger than $10 \%$ of the lattice volume after $N=6$. Our data clearly suggests that only smaller values of $N$ support the existence of macroscopic loops.

\subsubsection*{Poisson-Dirichlet behavior} Ref.\cite{thesis} proposed a polynomial behaviour for distribution of the longest loops in the system, characterized by a fugacity dependent functional form Eq.(\ref{plink_pd_eqn}). This behaviour is characterized by "shoulder-like" plateaus on a double logarithmic scale, which essentially arise from the power law $(f\mathcal{L}-l)^{\theta-1}$, where $f\mathcal{L}$ is the length of the longest loop and $\theta$ is the fugacity. We observe such plateaus with quick fall-offs upto $N=4$. For larger $N$, the distribution of the largest loops in the system appears to be governed by a different functional form, showing neither a sharp fall-off nor a flat regime characteristic of the Poisson-Dirichlet behaviour.

\begin{figure}[h!]
    \centering
    \includegraphics[width=1\columnwidth]{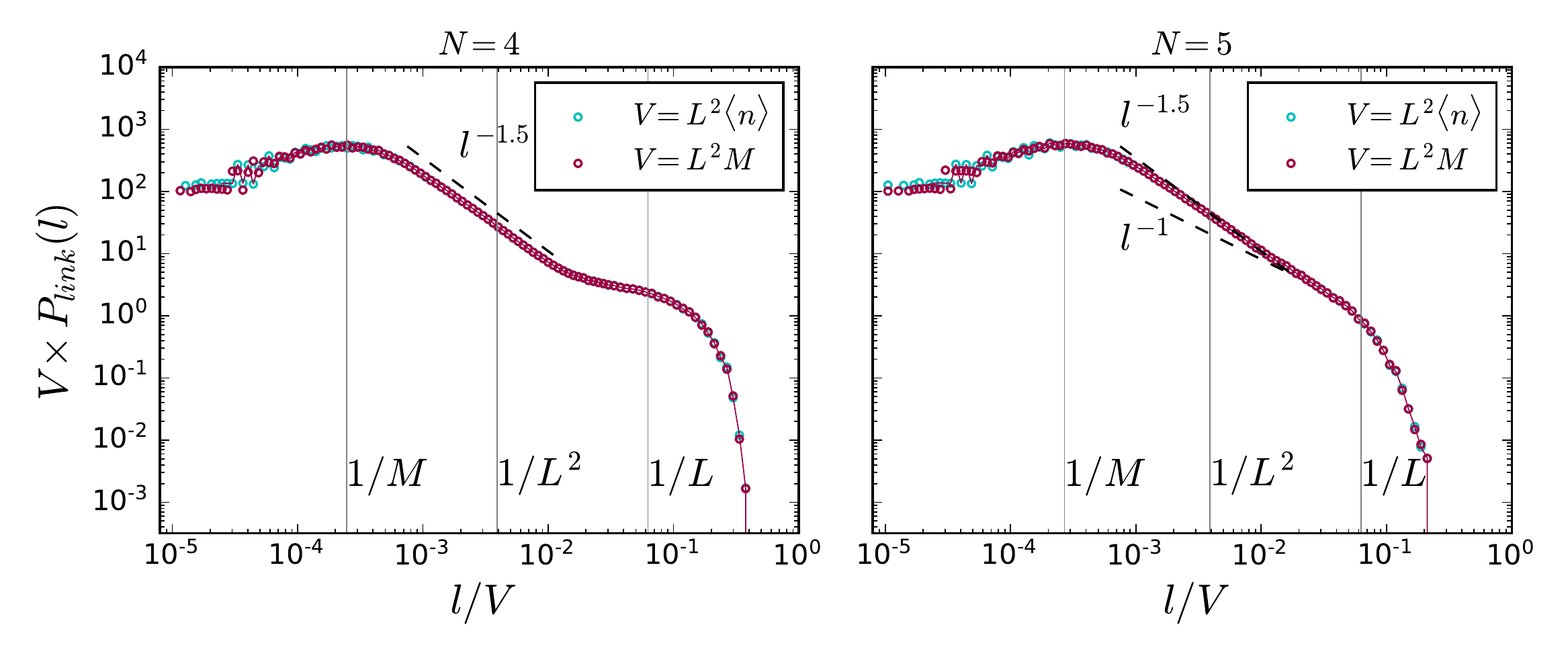}
    \caption{$V P_{\text{link}}(l)$ vs $l/V$ for $N=4$ and $N=5$, computed for SSE ensembles including (\textit{cyan}) and excluding (\textit{red}) identities}
    \label{N=45}
\end{figure}

\begin{figure}[h!]
    \centering
    \includegraphics[width=1\columnwidth]{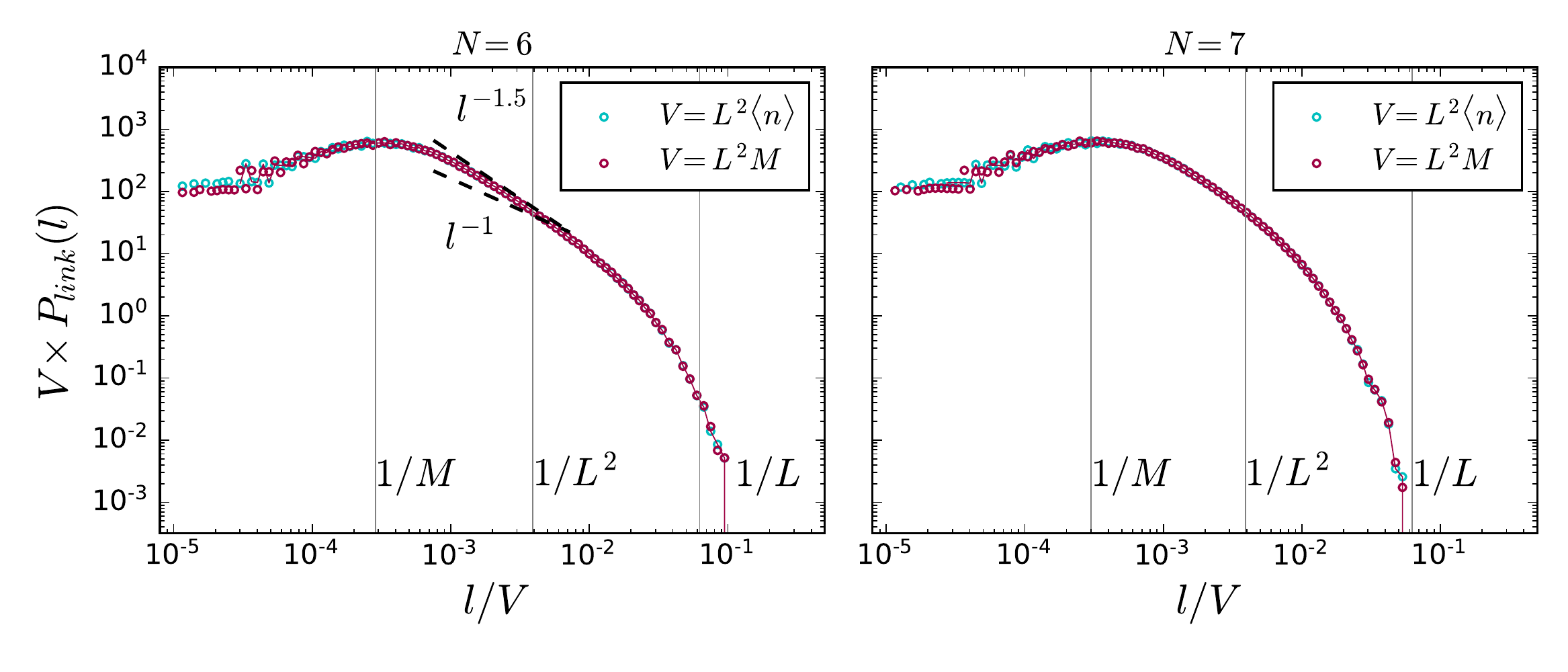}
    \caption{$V P_{\text{link}}(l)$ vs $l/V$ for $N=6$ and $N=7$, computed for SSE ensembles including (\textit{cyan}) and excluding (\textit{red}) identities}
    \label{N=67}
\end{figure}

\begin{figure}[h!]
    \centering
    \includegraphics[width=1\columnwidth]{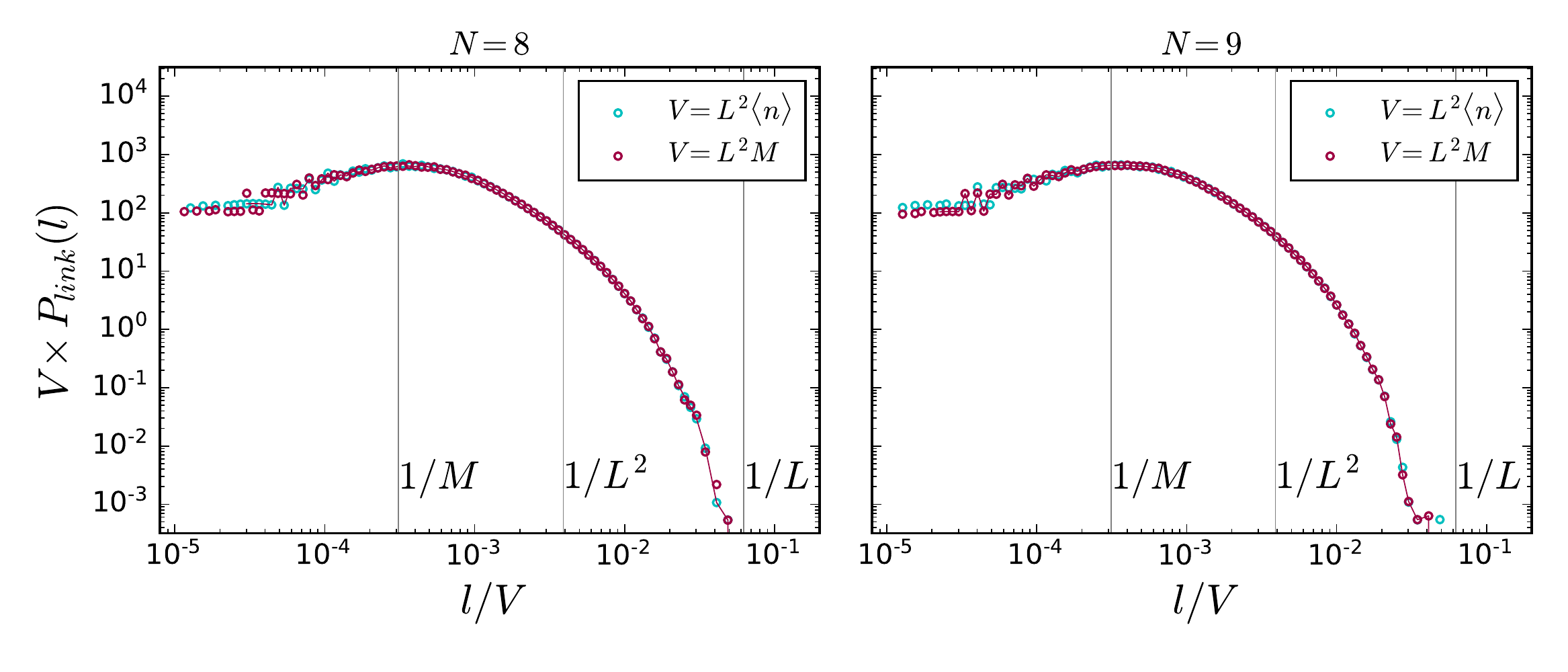}
    \caption{$V P_{\text{link}}(l)$ vs $l/V$ for $N=8$ and $N=9$, computed for SSE ensembles including (\textit{cyan}) and excluding (\textit{red}) identities}
    \label{N=89}
\end{figure}

\subsubsection*{Brownian character for short loops} The shorter loops in the models considered in Ref.\cite{thesis} showed Brownian character i.e. the distribution of loop lengths through a selected link followed the return-probability of random walker. In three dimensions this implies $P_{\text{link}}(l) \sim l^{-1.5}$. We observe that the loops longer than the size of the lattice $L^2 \sim O(1/M)$, but not long enough to be in the Poisson-Dirichlet regime, showcase a power-law fall-off with  Brownian exponent $1.5$. Similar to the lattice loop models, this power-law behaviour is present for smaller $N$s that lie in the extended phase. Interestingly, this behaviour also persists for larger $N$s and one can observe remnants of a linear region in the distribution. However, the exponent for $N>4$ starts deviating from the Brownian value.



\subsubsection*{Choice of SSE ensemble} 

For all values of $N$ considered, the loop lengths were calculated for two different choices of ensembles that one can use to generate loops from SSE. The data in \textit{red} is from an ensemble with fixed number of operators $M$ which allows for identity operators, and data in $cyan$ is from the ensemble with no identity operators where $n$ indicates the number of bond-operators. One must note that the latter ensemble is different from conventional loop models where the volume of the lattice is fixed, whereas here we have a fluctuating length $n$. However, in practice, $n$ varies typically as a Gaussian around a mean value and we expect both the ensembles to show similar qualitative behaviour apart from possible fluctuations. The data indicates that, for the systems with $L=\beta$ considered here, both the ensembles agree remarkably well for all $N$s with their distributions coinciding near perfectly apart from very small length scales.

\subsection{Ratio of moments}

\begin{figure}[h!]
    \centering
    \includegraphics[width=0.85\columnwidth]{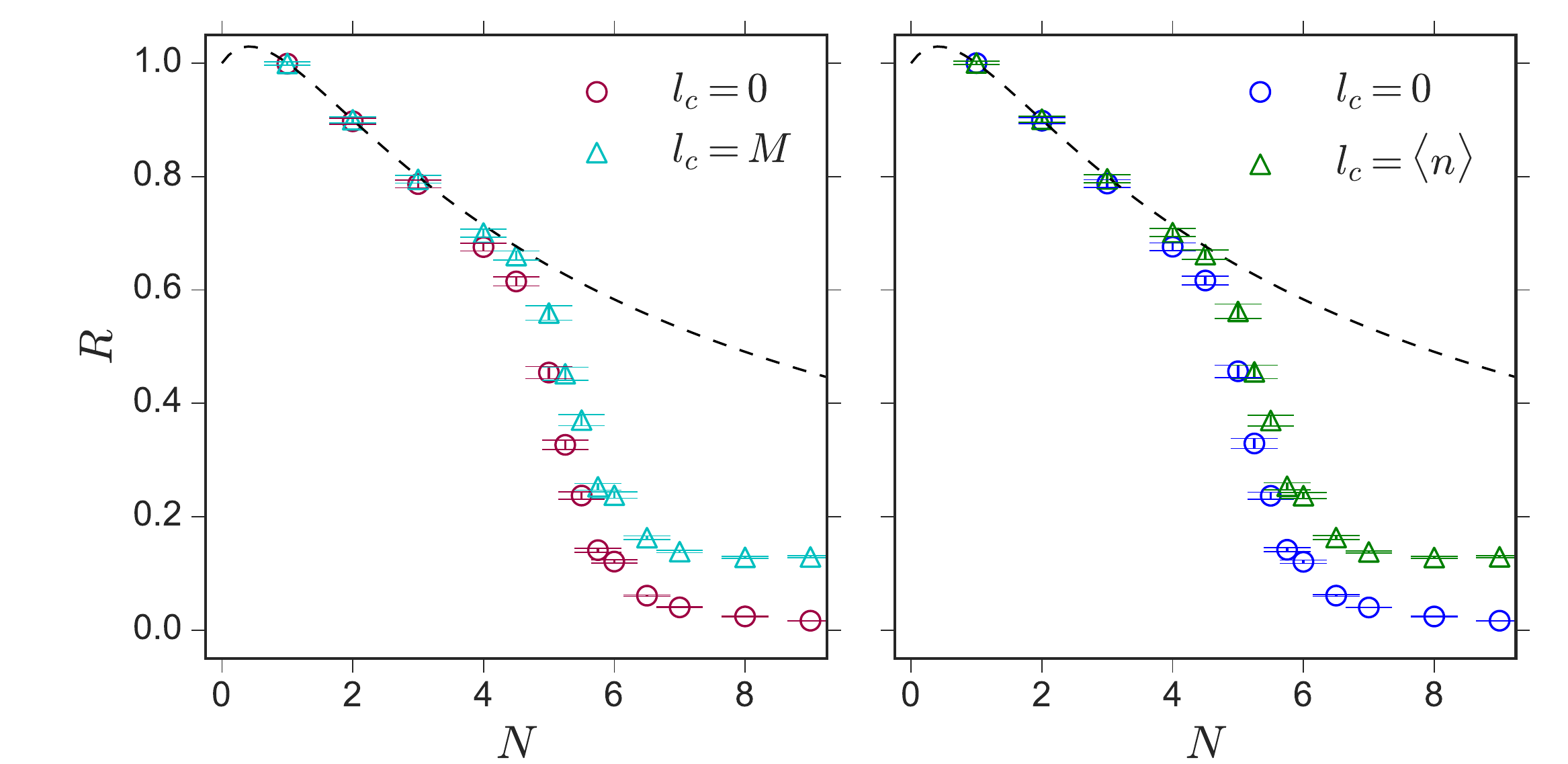}
    \caption{Ratio of moments of loop lengths $R$ vs $N$ considering cutoff lengths $l_c=0 \text{ and } V/L^2$ for SSE ensembles including  (\textit{left}) and excluding (\textit{right}) identities}
    \label{R_N}
\end{figure}  

The ratios of moments of the loop lengths offer  scale-independent quantities to study properties of the joint length distribution. Here, we compute the single-loop and two-loop moments defined in Chapter 2. The data points are collected by averaging over 10,000 Monte Carlo steps and the error bars are estimated using a Jackknife procedure. \cite{data}

We observe that for small values of $N$, the computed moments match the Poisson-Dirichlet estimate shown by the dashed curve. Particularly for the $N$s that exhibited the "shoulder-like" plateaus  for longest loops in $P_{\text{link}}(l)$, as is characteristic of the Poisson-Dirichlet behaviour. The ratios at $N=4$ show a small deviation from the estimate and for higher $N$, the system appears to transition to a different phase with the ratios saturating at a particular value. The described behaviour is seen for the single-loop moment ($R$) as well as the two-loop moments ($Q_2$ and $Q_3$).

\begin{figure}[h!]
    \centering
    \includegraphics[width=0.85\columnwidth]{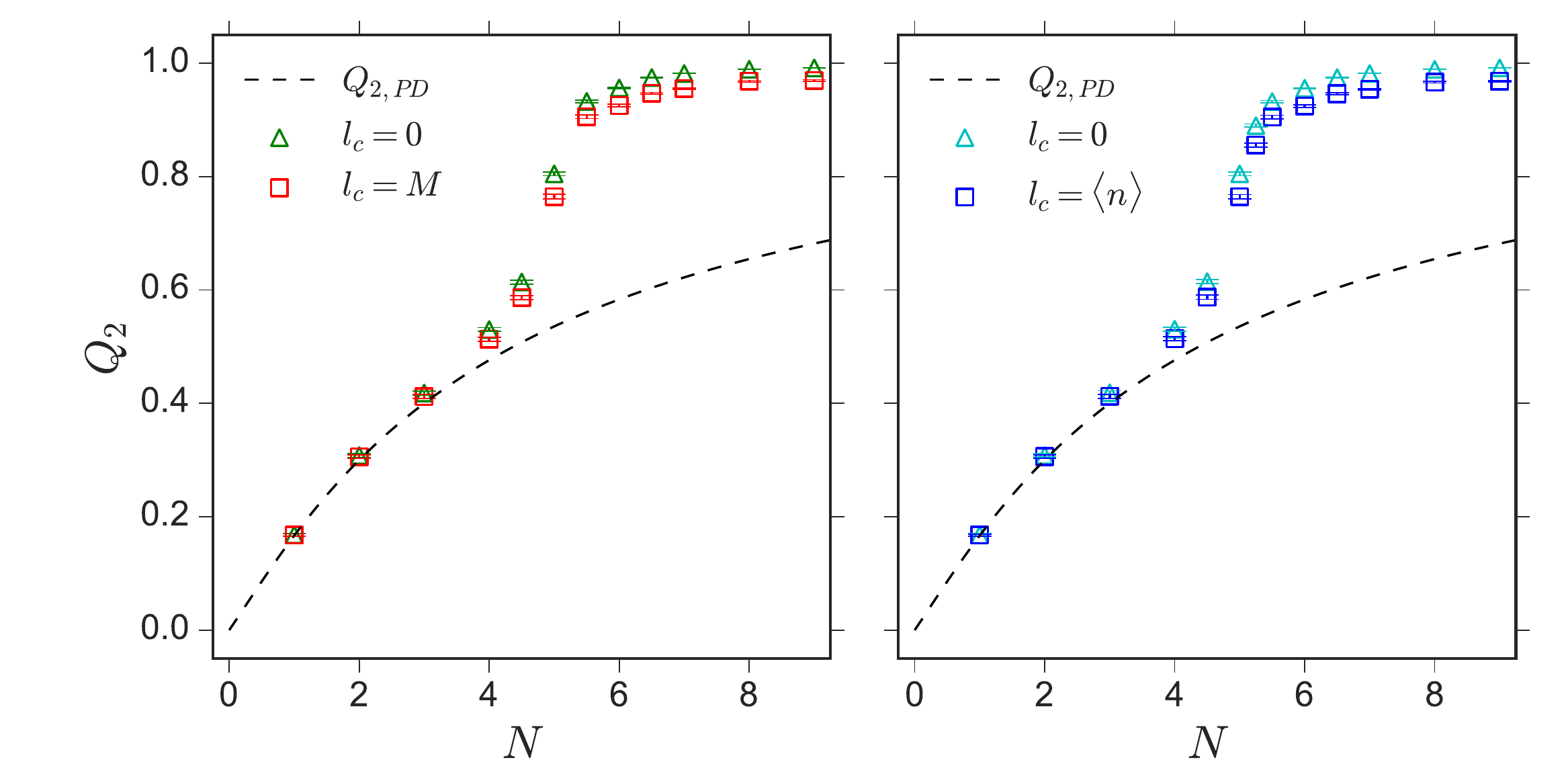}
    \caption{Ratio of moments of loop lengths $Q_2$ vs $N$ }
    \label{Q2_N}
\end{figure}

\begin{figure}[h!]
    \centering
    \includegraphics[width=0.85\columnwidth]{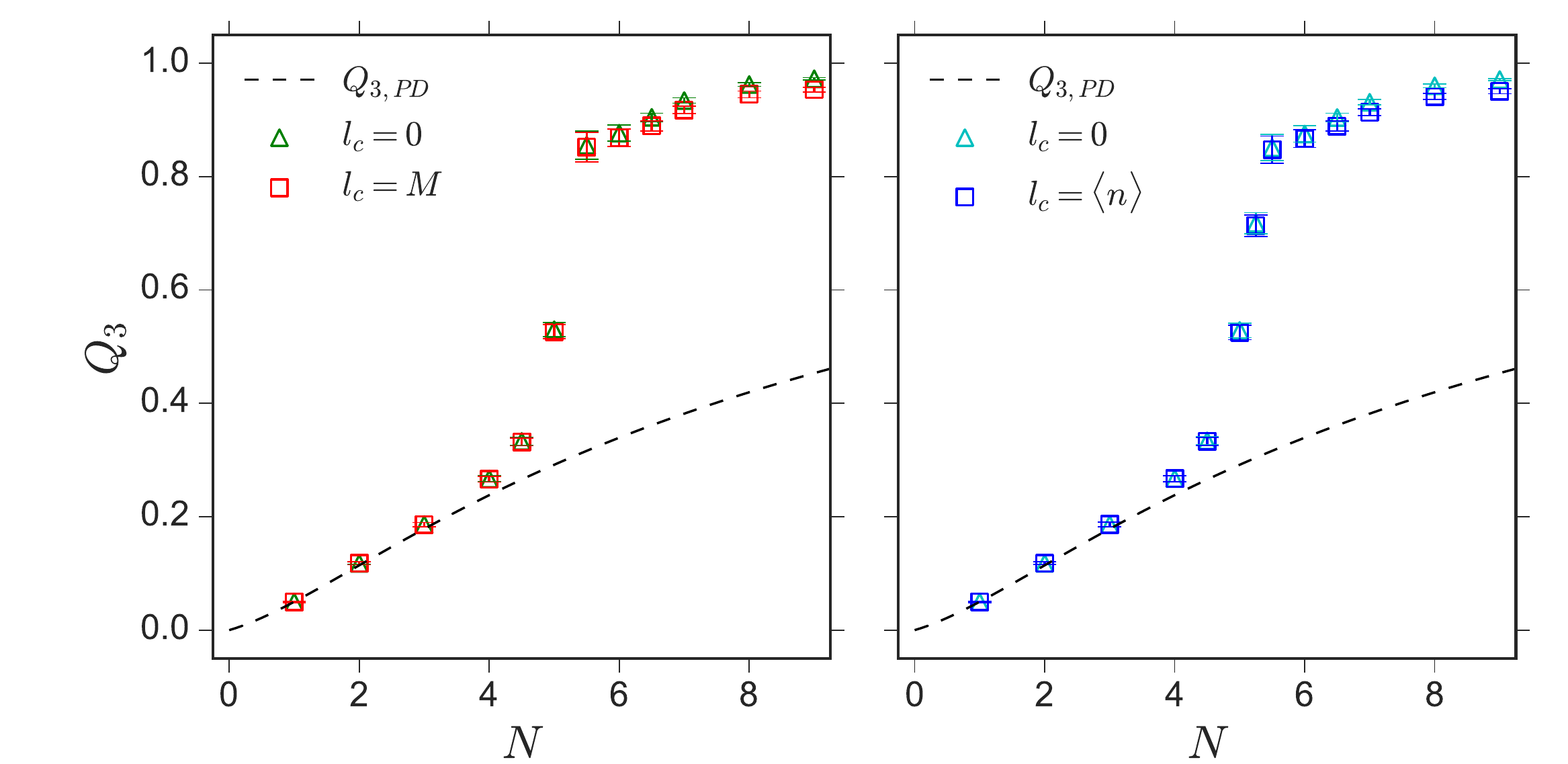}
    \caption{Ratio of moments of loop lengths $Q_3$ vs $N$}
    \label{Q3_N}
\end{figure}

In every figure, we compare the ratios calculated using lengths of all the loops in the ensemble ($\circ$) and for loops longer than a cutoff ($\Delta$). We expect the latter to be closer to the Poisson-Dirichlet estimate, which is strictly speaking valid for the longest loops. The ratios $R$ and $Q_2$ show this behaviour. We also study the dependence on the choice of SSE ensemble by comparing values obtained for the series-expansion truncated at some $n_{\text{max}}=M$ (\textit{left panel}) to the case with varying $n$ $(\textit{right panel})$. We see identical behaviour for both ensembles as is expected due to the good overlap seen in the distribution $P_{\text{link}}(l)$.

\subsection{Fitting the distribution for longest loops}

\begin{figure}[h!]
    \centering
    \includegraphics[width=0.6\columnwidth]{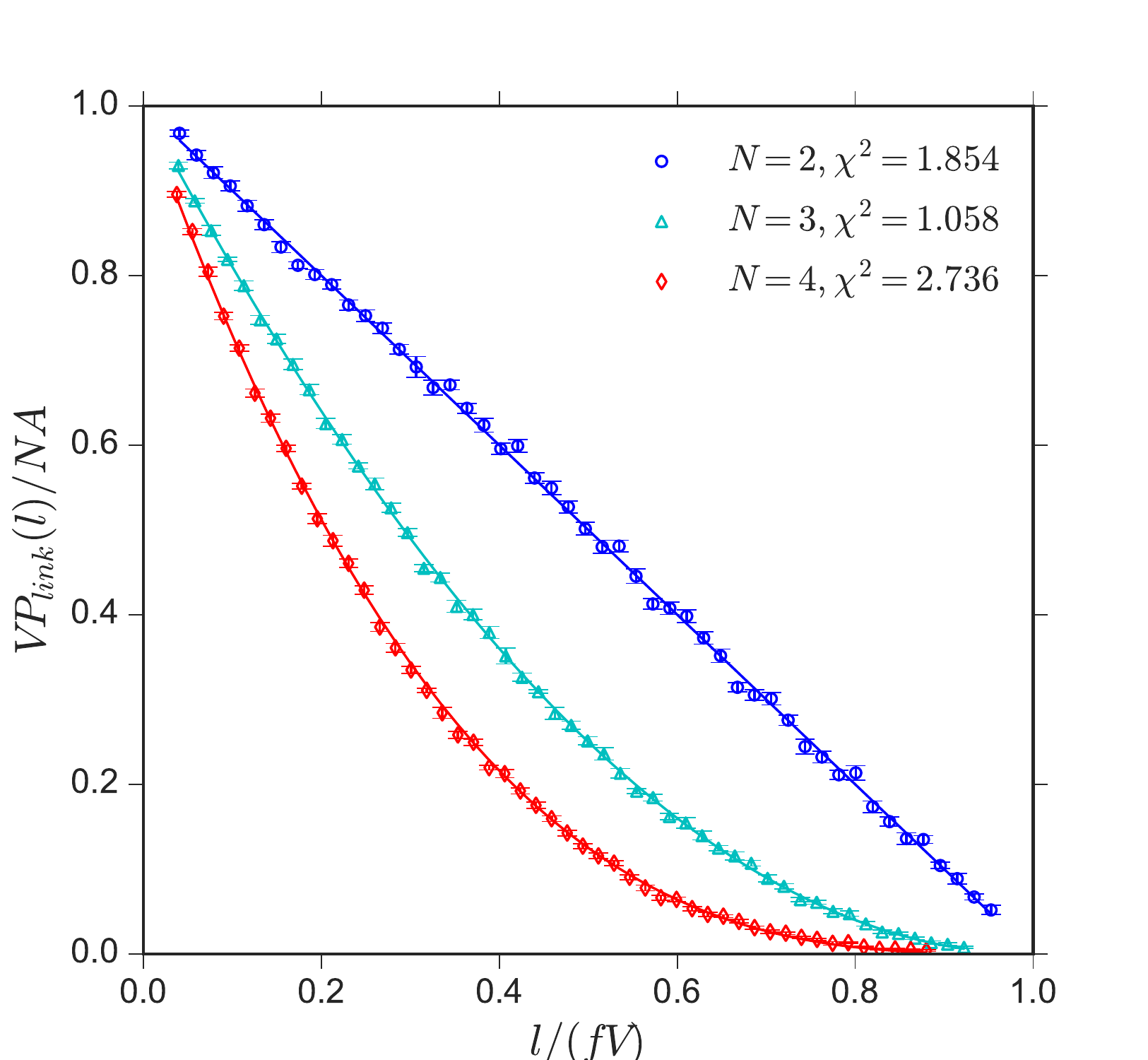}
    \caption{ Fitting a polynomial Eq.(\ref{plink_pd_eqn}) using $\chi^2$ estimation to loop length distribution $P_{\text{link}}(l)$ for $N=2,3$ amd $4$. }
    \label{pdfit}
\end{figure} 

\subsubsection*{Poisson-Dirichlet regime}
One of the motivations for our work is to check the validity of Poisson-Dirichlet behaviour in the extended phase. For this purpose we investigate whether the functional form Eq.(\ref{plink_pd_eqn}) for distribution in the PD regime of lattice-loop models describes the shoulder-like plateaus we have observed in $P_{\text{link}}(l)$ for $N=2,3$ and $4$. We check the fit of the following function
\begin{equation}
    A\left (1 - \frac{l}{fV} \right )^{N-1},
\end{equation}
to our data, where $A$ and $f$ are fitting parameters. We observe that distribution for $N=2,3$ and $4$ does indeed match the polynomial behaviour expected by Eq.(\ref{plink_pd_eqn}), although with sub-optimal fitting error $(\chi^2 >1)$. The larger values of $N$, not included here, have significant error with $\chi^2 >10$ for $N=5$, and $\chi^2 \to \inf$ when $N>5$. We infer that the length distribution of the longest loops for these values of N cannot be governed by the Poisson-Dirichlet behaviour. 

\subsubsection*{Non Poisson-Dirichlet regime (Exponential fit)}

\begin{figure}[h!]
    \centering
    \includegraphics[width=0.6\columnwidth]{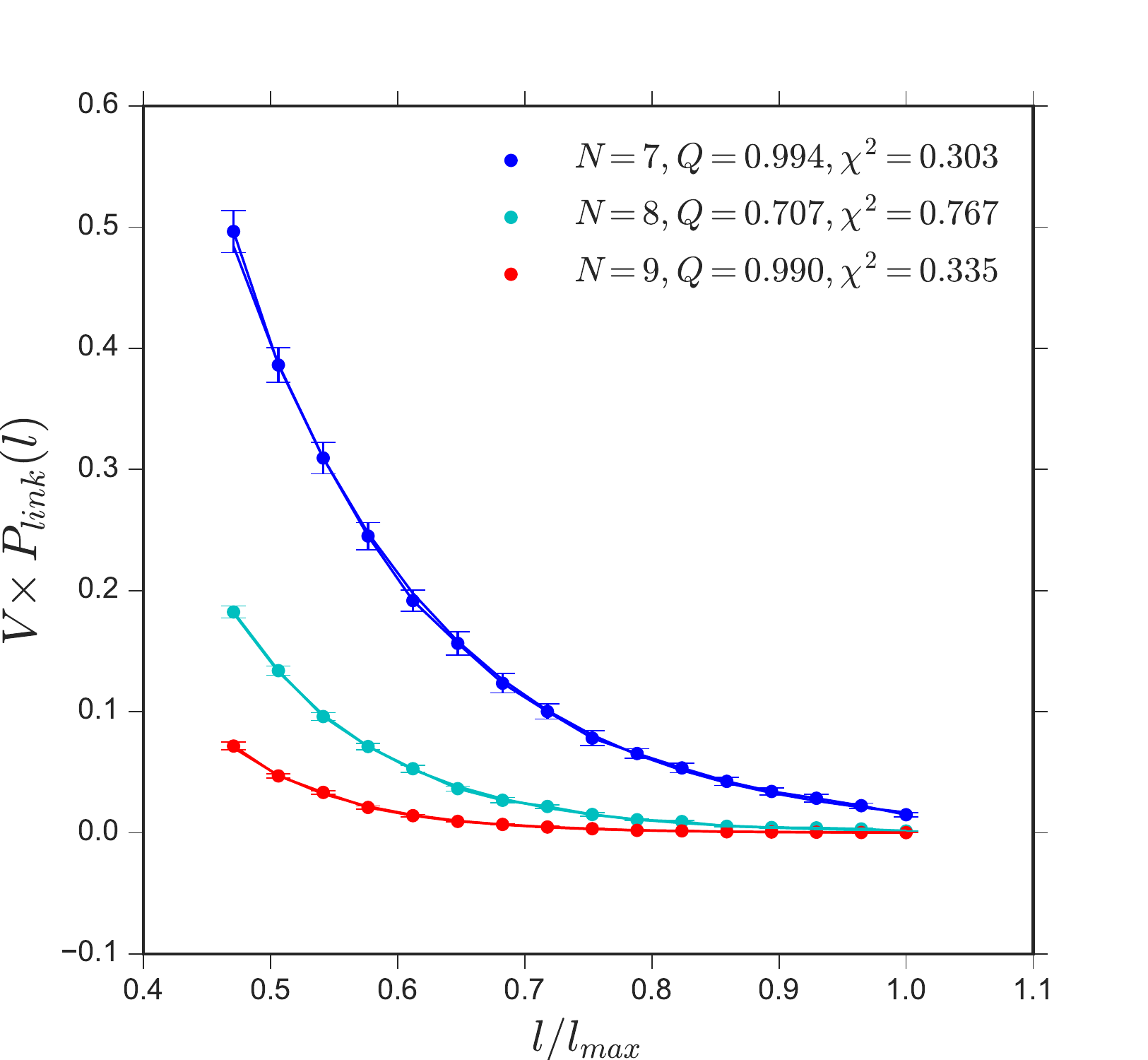}
    \caption{Fitting an exponential decay Eq.(\ref{exp_fit}) using $\chi^2$ estimation to loop length distribution $P_{\text{link}}(l)$ for $N=7,8$ and $9$}
    \label{expfit_789}
\end{figure}

\begin{figure}[h!]
    \centering
    \includegraphics[width=0.6\columnwidth]{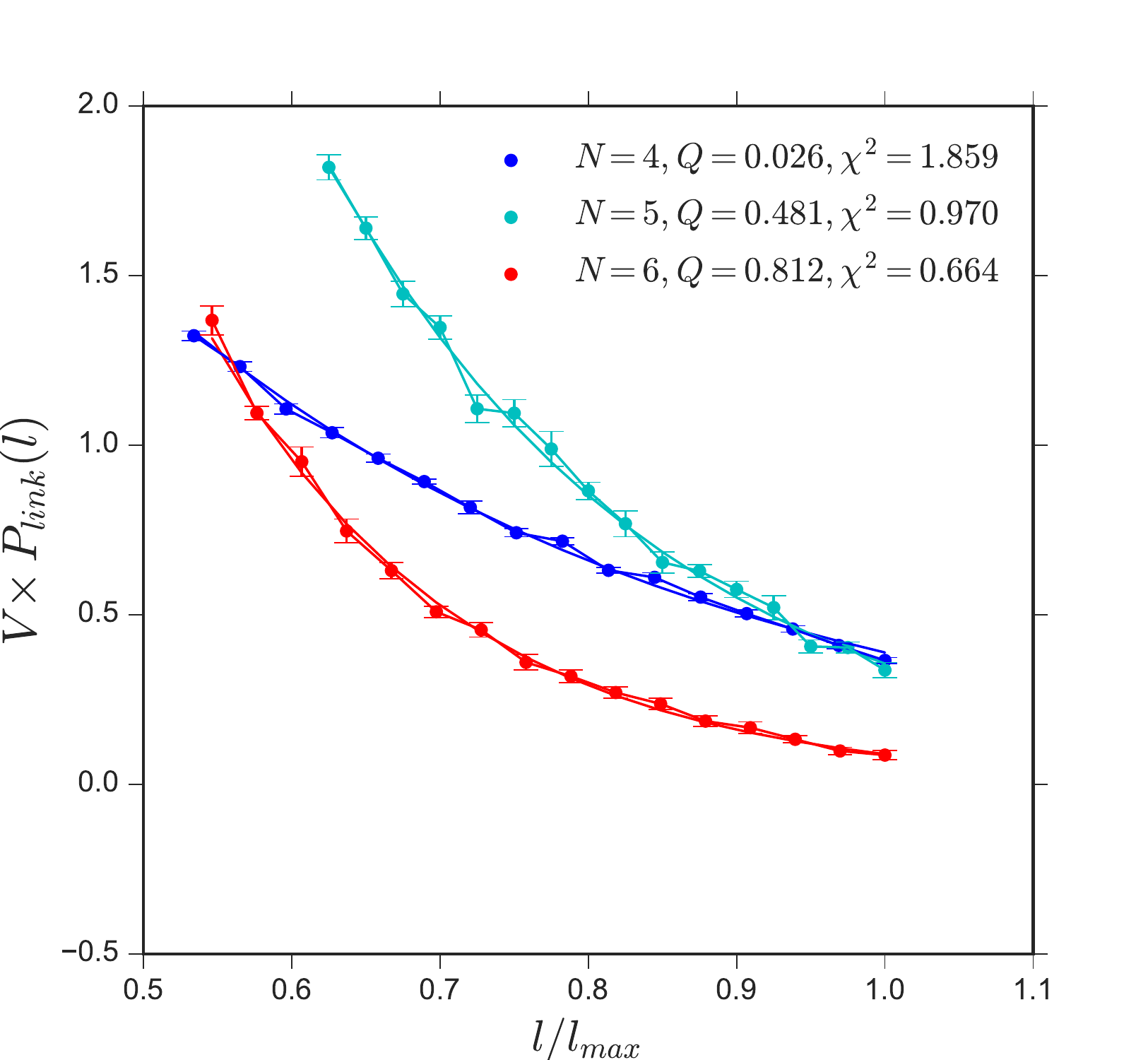}
    \caption{Fitting an exponential decay Eq.(\ref{exp_fit}) using $\chi^2$ estimation to loop length distribution $P_{\text{link}}(l)$ for $N=4,5$ and $6$}
    \label{expfit_456}
\end{figure}

For $N=5$ to $9$, we observed a fall-off in $P_{\text{link}}(l)$ that is faster than the linear behaviour of Brownian loops, but slower than that in PD regime. We thus attempt to fit the data to an exponential decay,
\begin{equation}
    K \exp(-t l),
    \label{exp_fit}
\end{equation} 
with $K$ and $t$ as the fitting parameters. The lengths considered for fitting are chosen so as to exclude any linear region $ \sim l^{-\alpha}$, if present, to the left, and largest bins in distribution that have zero entries, to the right. The results of the fitting are thus presented with $l/l_{\text{max}}$ in the horizontal axis for clarity. Here $l_{\text{max}}$ is the largest length considered.

For $N=7,8$ and $9$, we observe that the data fits the exponential decay with high quality factor $Q>0.7$ and reasonable fitting error $\chi^2 < 1$. As per our analysis of the ratios of moments the largest values $N=8$ and $9$ appear to saturate to a fixed value. We assume that these values of $N$ serve as good candidates to explore the distribution that characterizes the short-loop phase. It is noteworthy that the data for $N=7$ and 9 fit the exponential decay remarkably well with $Q>0.99$.

\newpage

\section{Tuning $\beta$}
\label{tbeta}

As the inverse temperature $\beta= 1/T$ is decreased, the system transitions from a low temperature extended phase to a high temperature disordered phase. Similar to Sec.(\ref{Ntx}), we study below how the length distribution of loops passing through a randomly selected link $P_{\text{link}}(l)$, and the ratio of moments of the joint length distribution $R, Q_m$ vary across the transition by looking at selected values of $\beta$. 
The simulations are carried out for a 2D square lattice of side length $L=16$ for a spin-$1/2$ system i.e. $N=2$, to reproduce the familiar extended phase for large $\beta$. The distribution is sampled using $10^5$ Monte Carlo steps.

\subsection{Loop length distribution}

We consider six values of $\beta$ to illustrate behaviour across the transition. The key aspects of the distribution for consideration are as follows:

\subsubsection*{Peaks at $l = k \times M$ in the disordered-phase}

\begin{figure}[h!]
    \centering
    \includegraphics[width=\columnwidth]{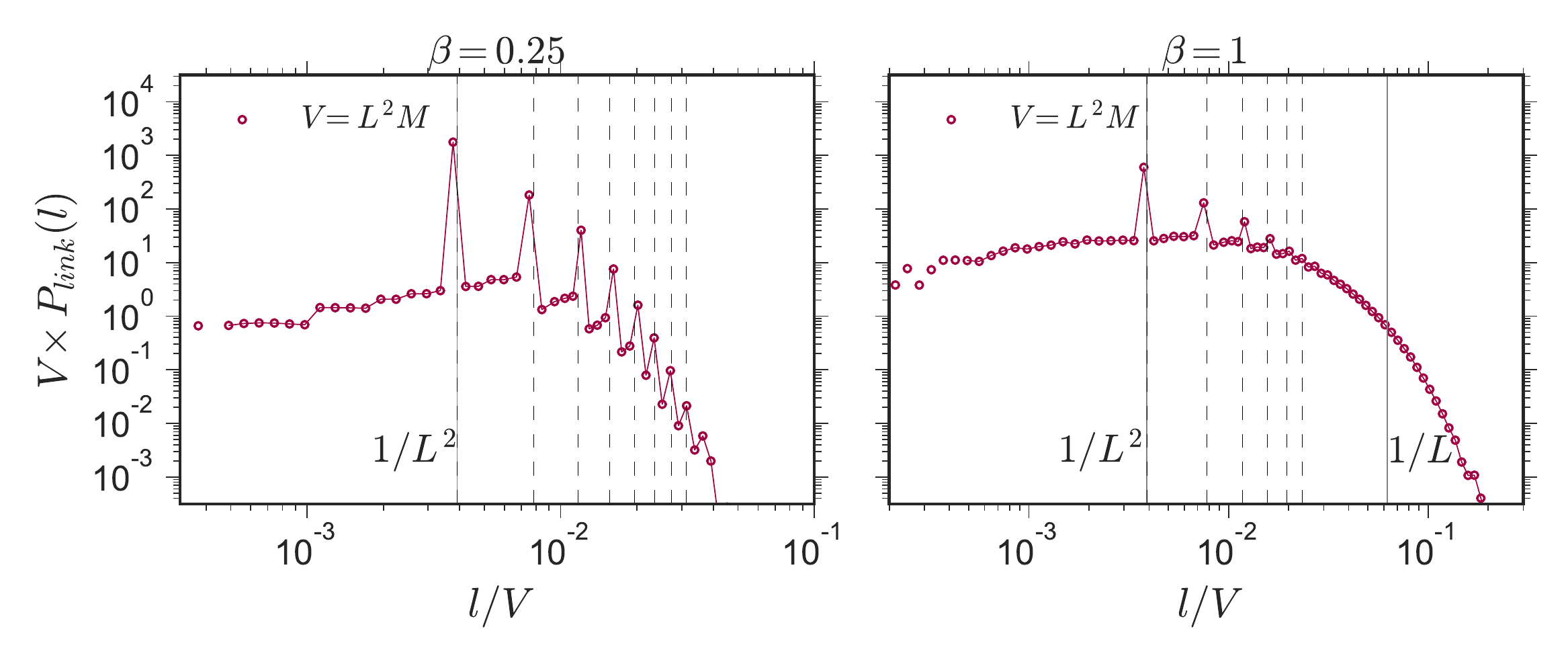}
    \caption{$V P_{\text{link}}(l)$ vs $l/V$ for $\beta=0.25$ and $\beta=1$, for $N=2$, indicating distribution in the high-temperature/disordered phase. Here we consider only the SSE ensembles including identities.}
    \label{beta_0.25_1}
\end{figure}

In the disordered-phase (e.g. Fig. (\ref{beta_0.25_1})), we see the presence of sharp peaks in the distribution for loop-lengths $l = k \times M$, i.e at integral multiples of length along the "time-axis" M. Since the distribution is plotted using a double logarithmic scale, the presence of the peaks indicates that an overwhelming majority of loops in the ensemble have the special values of loop-lengths. Loops with these special lengths can be understood to emerge from a simple geometric constructions on the space-time lattice. The first of these lengths, namely $l=M$, denotes vertical/one-dimensional strands which stay on a particular site and loop back due to the time-periodic boundary conditions. When we consider two such strands on neighboring sites and include an operator on the corresponding bond between the sites, we get a loop of length $l=2M$. Proceeding in this fashion, one can construct loops of length = $l=k \times M$ when a sequence of $k$ sites is chosen where each site is adjacent to the previous one, and correspondingly $k$ operators are present at non-adjacent times. The abundance of loops of this kind in the disordered-phase indicates that the operators are very sparsely populated across sites i.e. the expected value of number of operators per site is very low ($ < 1$). One expects the peaks at $l=k\times M$ to predominantly govern the physics of the disordered phase. 

\subsubsection*{Towards the low temperature phase}

\begin{figure}[h!]

    \centering
    \includegraphics[width=\columnwidth]{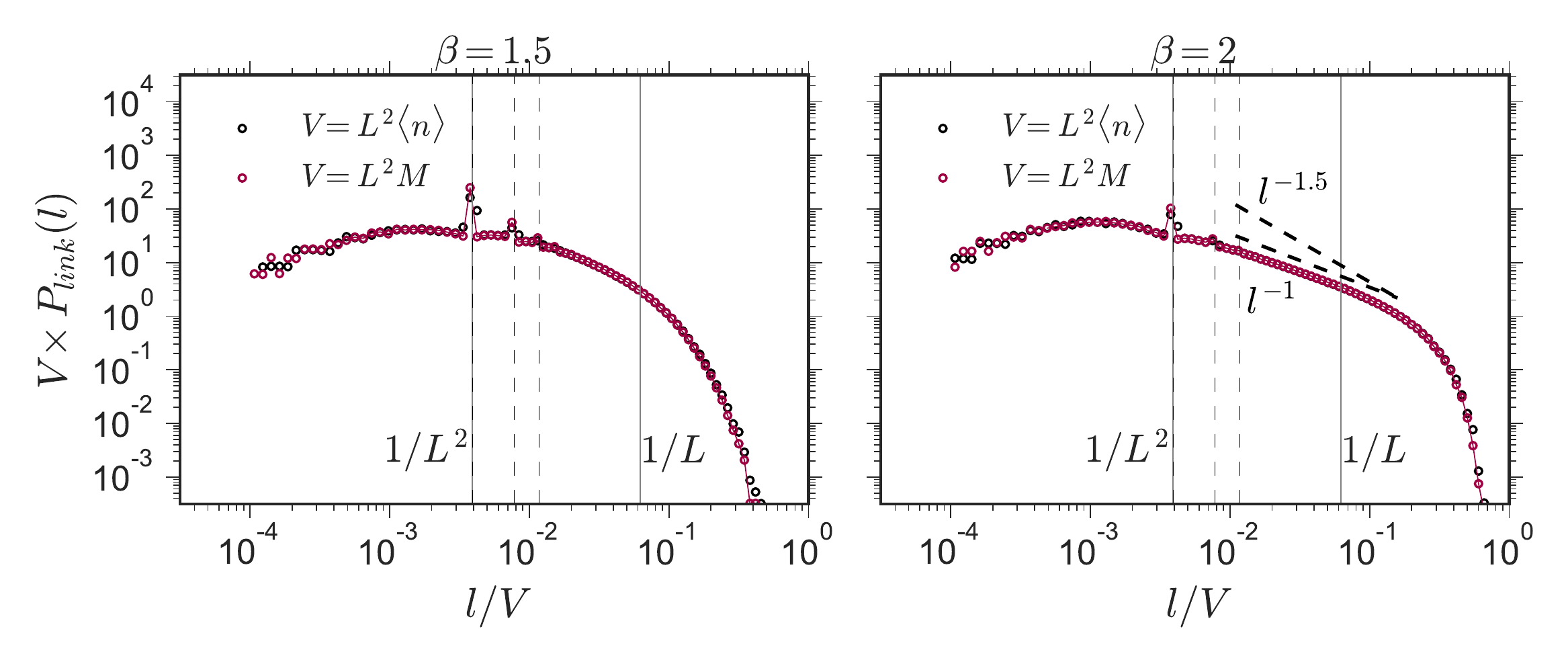}
    \caption{$V P_{\text{link}}(l)$ vs $l/V$ for $\beta=1.5$ and $\beta=2$, for $N=2$, computed for SSE ensembles including (\textit{cyan}) and excluding (\textit{red}) identities}
    \label{beta_1.5_2}
\end{figure}

\begin{figure}[h!]
    \centering
    \includegraphics[width=\columnwidth]{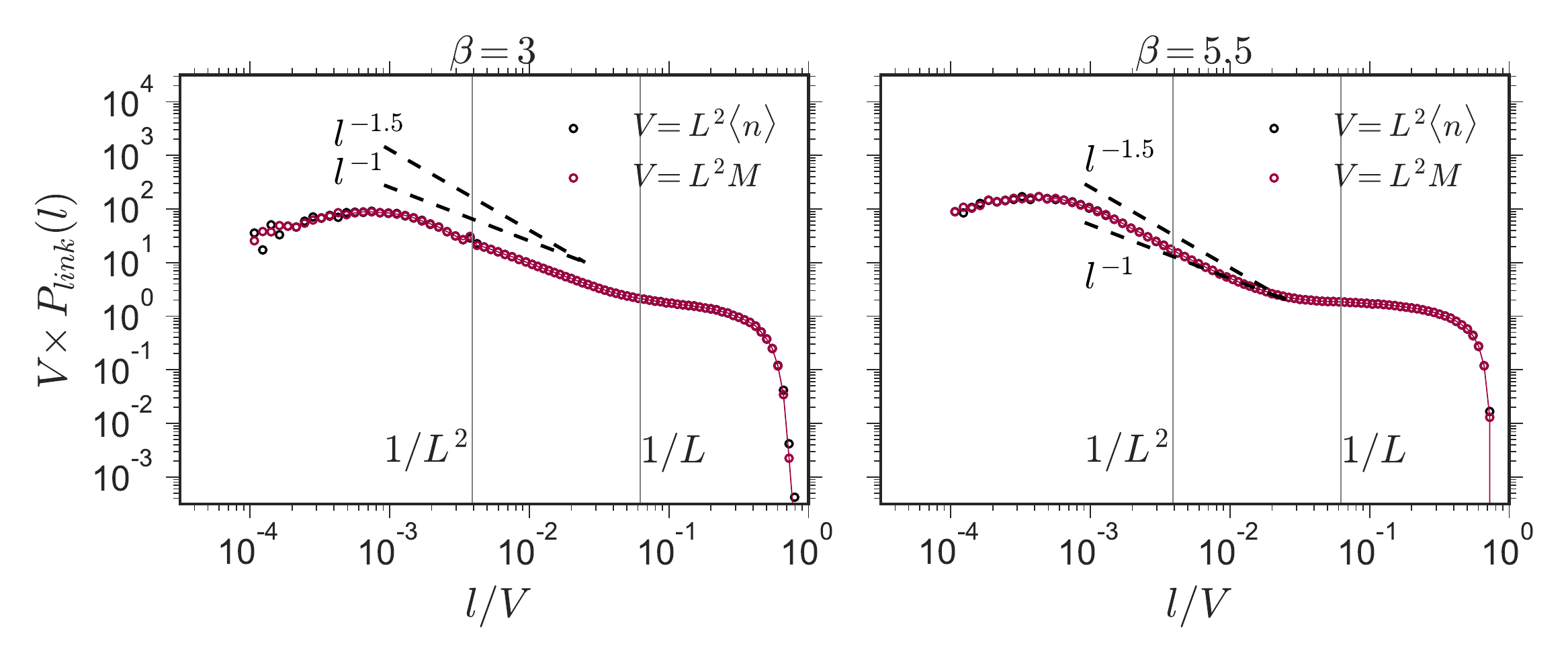}
    \caption{$V P_{\text{link}}(l)$ vs $l/V$ for $\beta=3$ and $\beta=5.5$, for $N=2$, computed for SSE ensembles including (\textit{cyan}) and excluding (\textit{red}) identities}
    \label{beta_3.5_5}
\end{figure}

As the temperature is lowered, i.e. parameter $\beta$ is increased in the simulations, we expect the system to start showing signs of the characteristics of Poisson-Dirichlet distribution that we know is applicable for $\beta=L$. First and foremost, as $\beta$ is increased, the prevalence of peaks at the special loop lengths decreases. The figures Fig.(\ref{beta_0.25_1}), Fig.(\ref{beta_1.5_2}) and Fig.(\ref{beta_3.5_5}) show a decline, for rising $\beta$ in both the number of peaks and their amplitudes
Secondly, in our plot for $\beta=1.5$, we start seeing a power-law decay, characteristic of Brownian loops, emerging for intermediate lengths. For $\beta=2$ and 3, we see an extended power-law decay region but the fall-off is closer to $l^{-1}$ than to the Brownian $l^{-1.5}$. For $\beta=5.5$, we see a steeper decay $l^{-d}$ with power $ 1<d<1.5$ but still not completely Brownian. Lastly, in Fig.(\ref{beta_3.5_5}), the largest $\beta$ that we consider, demonstrate the shoulder-like plateaus characteristic of Poisson-Dirichlet distribution. Therefore, already for $\beta=5.5$, we see the signs of the Brownian power-law decay and shoulder-like Poisson-Dirichlet plateaus that we know to be applicable in the low temperature phase.   

\subsubsection*{Choice of SSE ensemble}

In Fig. (\ref{beta_1.5_2}) and Fig.(\ref{beta_3.5_5}), the loop lengths are calculated using both the ensembles that one can use to generate loops from SSE. The data in \textit{red} is from an ensemble with fixed number of operators $M$ which allows for identity operators, and data in \textit{black} is from the ensemble with no identity operators where $n$ indicates the number of bond-operators. The data indicates good agreement between both ensembles and the corresponding distributions coinciding well across wide range of length scales. The distribution around the peaks $k=l \times M$, prevalent in the disordered phase spreads out in the ensemble without identities. Consequently, we see wider and shorter peaks, see Fig.(\ref{beta_1.5_2}). The ensemble excluding identities is not shown in Fig.(\ref{beta_0.25_1}) for clarity. 







  


 \chapter{Conclusion \& Outlook}

One of our motivations starting out was to verify the applicability of Poisson-Dirichlet behaviour in the extended phase. In this pursuit, we analyzed the distribution of loops through a link $P_{\text{link}}(l)$ and the moments of the joint-length distribution. For $N=2,3$ and $4$, we infer that, akin to classical 3D loop models with a fugacity, 
\begin{itemize}
    \item there exist loops of macroscopic length scales
    \item the single-loop ($R$) and joint-loop ($Q_2,Q_3,Q_4$) moments, along with plateaus in $P_{\text{link}}(l)$ indicate that macroscopic loops are governed by the Poisson-Dirichlet distribution
    \item there are loops shorter than the PD regime, for which the $P_{\text{link}}(l)$ decays as per Brownian return probability $l^{-d/2}$
\end{itemize}
For larger values of $N$, 
\begin{itemize}
    \item the system transitions out of regime of applicability of PD behaviour for $N=4$ to 7, with the moments saturating at $N=8,9$
    \item the $P_{\text{link}}(l)$ decays exponentially with $l$ for $N=8,9$, similar to decay of correlations in valence bond solids 
\end{itemize}

In Sec.(\ref{tbeta}), we studied the transition from low-temperature phase to the high-temperature/disordered phase. We learn that:
\begin{itemize}
    \item In the disordered phase, the ensemble is largely dominated by loops of special lengths $l = k \times M$. This arises from the fact that operators are sparsely populated in time. 
    \item The distribution of loop-lengths in the high-temperature phase and in the low-temperature large N phase are both different, in different ways, from the low N low temperature phase that exhibits macroscopic loops and is governed by power-law Brownian decay and Poisson-Dirichlet plateaus. The high-temperature N=2 phase is dominated by loops of certain lengths, whereas, the low-temperature large N distribution is smooth but of a different qualitative nature, namely exponential decay. 
\end{itemize}

Going forward, we would like to explore the origin of the exponential decay in $P_{\text{link}}(l)$ of loop-lengths for large $N$s. In this regard, one can explore the mathematical relationship between loop-lengths and correlations. Another approach is to compute the distribution  starting from the first-principles akin to the work of D.Ueltschi for the extended phase \cite{ueltjmp}. Finally, we are curious about further insights one can gain into the physics of SU(N) antiferromagnets using a loop-ensemble description.

 \phantomsection

 \addcontentsline{toc}{chapter}{Bibliography}


\appendix
\chapter{Details of field theory calculations for Chapter 2}

\section{Relation to $O(3)$ universality class}

In the main text, we saw that the continuum description conjectured by Nahum et.al., based on the appropriate symmetries, is the $CP^{n-1}$ $\sigma$-model with field $Q$
\begin{equation}
    \mathcal{L} = \frac{1}{2g} \text{tr}(\nabla Q)^2.
\end{equation}
The traceless Hermitian $n \times n$ matrix $Q$ also satisfies the constraint that arises from the normalization of $\mathbf{z}$, 
\begin{equation}
    (Q+1)^2 = n(Q+1).
\end{equation}
One can also consider a \textit{soft-spin formulation} where $Q$ can be taken to be an arbitrary traceless Hermitian matrix,
\begin{equation}
    \mathcal{L}_{\text{soft}} = \text{tr}(\nabla Q^2) + t \text{tr}(Q^2) + g \text{tr}(Q^3) + \lambda \text{tr}(Q^4) + \lambda^{'} (\text{tr}(Q^2))^2.
    \label{softspin}
\end{equation}

Next, we will explore how the special case of $n=2$ i.e. the $CP^{1}$ model is equivalent to the $O(3)$ model. Note that the field in this case is a $2 \times 2$ traceless Hermitian matrix, which can be parametrized in terms of the Pauli matrices $\sigma^i$, as
\begin{equation}
    Q = \frac{1}{\sqrt{2}} \sigma^i S^i, 
\end{equation}
where summation over the repeated index is implied. One can plug this term-wise into Eq.(\ref{softspin}) and get terms of the form,
\begin{equation}
    \text{tr}(Q^2) = \text{tr} \frac{1}{2} \sigma^i \sigma^j S^i S^j, \quad 
    \text{tr}(\nabla Q^2) = \text{tr} \frac{1}{2} \sigma^i \sigma^j \nabla S^i \nabla S^j, \quad
    \\
    \text{tr}(Q^3) = \text{tr} \frac{1}{2\sqrt{2}} \sigma^i \sigma^j \sigma^k S^i S^j S^k, 
\end{equation}
and so on. Using the anticommutation property of Pauli matrices $\{\sigma^i,\sigma^j\} = 2 \delta^{ij}$, the terms with even powers of $Q$ become
\begin{equation}
    \text{tr}(Q^2) = S^2, \quad
    \text{tr}(\nabla Q^2) = (\nabla S)^2, \quad 
    (\text{tr}(Q^2))^2 = (S^2)^2.
\end{equation}
Using a combination of anitcommutation relations with the cyclicity of trace, one can also show that, 
\begin{equation}
    \text{tr}(Q^3) = 0 , \quad
    \text{tr}( Q^4) = ((S)^2)^2/2.
\end{equation}
Hence Eq.(\ref{softspin}) becomes, 
\begin{equation}
    \mathcal{L}_{\text{soft}} = (\nabla S)^2 + tS^2 + u S^4,
\end{equation}
with $u=\lambda^{'} + \lambda/2$. The description for $n=2$ could be cast in terms three dimensional real vectors $S$ and thus one expects any transitions in the $n=2$ models to lie in the $O(3)$ universality class, see for e.g. \cite{prl2}.

\section{Correlations \& geometric observables}

The correlations of elements of the field $Q$ can be shown to be proportional to various geometrically defined events in the loop model. For instance, in a continuum notation one can define $G_k(x,y)$ to be the probability that $k$ different strands of loop pass through the points $x$ and $y$. Note that in a model were all loops close, $k$ must necessarily be even for $G_k$ to not vanish. In the four-coordinated lattice models we discussed, $G_2(l_1,l_2)$ is the probability for the two links $l_1$ and $l_2$ to lie on a single loop, and $G_4(x_1,x_2)$ is the probability that two nodes $(x_1,x_2)$ are connected by four strands. The distinction between nodes and links can be overlooked in the continuum model.

The relation of the probability of such geometrically defined events to the correlation functions can be understood as follows. Consider the the operator $Q^{\mu \nu}_l = z^{\mu}_l \bar{z}^{\nu}_l$, with $\mu \neq \nu$ at the link $l$. The expectation value for this operator becomes,
\begin{equation}
        \langle Q^{\mu \nu}_l \rangle = 
        \sum_C W_C \prod _{\text{loops in C}} \text{Tr } (\mathbf{z}_1^{\dagger} \mathbf{z}_2 )(\mathbf{z}_2^{\dagger} \mathbf{z}_3 )...(\mathbf{z}_l^{\dagger} \mathbf{z}_1) Q^{\mu \nu}_l,
    \label{Sfull2}
\end{equation}
The term inside the trace can be expanded as,
\begin{equation}
  \text{Tr} \sum_{\alpha_1,\alpha_2,..\alpha_l} 
  (\bar{z}_1^{\alpha_1} z_2^{\alpha_1} )
  (\bar{z}_3^{\alpha_2} z_3^{\alpha_2} )
  ..
  (\bar{z}_l^{\alpha_l} z_1^{\alpha_l} )
  Q^{\mu \nu}_l
  =  \sum_{\alpha_1,\alpha_2,..\alpha_l} 
 \delta^{\alpha_1,\alpha_l} \delta^{\alpha_1,\alpha_2}...
 \text{Tr } z_l^{\alpha_{l-1}} \bar{z}_l^{\alpha_l} 
 z^{\mu}_l \bar{z}^{\nu}_l 
 \label{megatrc}
\end{equation}
where the essential term on the link $l$ becomes,
\begin{equation}
    \text{Tr } z_l^{\alpha_{l-1}} \bar{z}_l^{\alpha_l} 
 z^{\mu}_l \bar{z}^{\nu}_l = A \delta^{\alpha_{l-1}, \nu} \delta^{\alpha_{l}, \mu}, \quad \mu \neq \nu
\end{equation}
where $A = n/(n+1)$ is just $\text{Tr } |z^{\mu}|^2 |z^{\nu}|^2$ for distinct $(\mu,\nu)$ and $A=\text{Tr } |z^{\mu}|^4$ if $\mu=\nu$.

Combining the above expression with the Kronecker $\delta$'s in Eq.(\ref{megatrc}), we see that $\langle Q^{\mu \nu}_l \rangle$ will always vanish for $\mu \neq \nu$. This is to be expected because earlier in the calculation of the partition function $Z$, we concluded that every loop in a configuration is left with a single color index, whereas $Q^{\mu \nu}_l = z^{\mu}_l \bar{z}^{\nu}_l$ describes a process where an incoming strand of color $\mu$ changes at link $l$ to become an outgoing strand of different colour $\nu$. However, this formulation is useful because one can calculate probabilities of specific geometric events. For example, the correlator $ \langle Q^{12}_l Q^{21}_{l'} \rangle$ sums over all configurations in which $l$ and $l'$ are joined by a loop with one arm of colour 1 and another of colour 2. Thus the geometric observable $G_2$ becomes,
\begin{equation}
    G_2(l,l') = \frac{n}{A^2} \langle Q^{12}_l Q^{21}_{l'} \rangle
\end{equation}

\begin{figure}
    \centering
    \includegraphics[width=0.9\columnwidth]{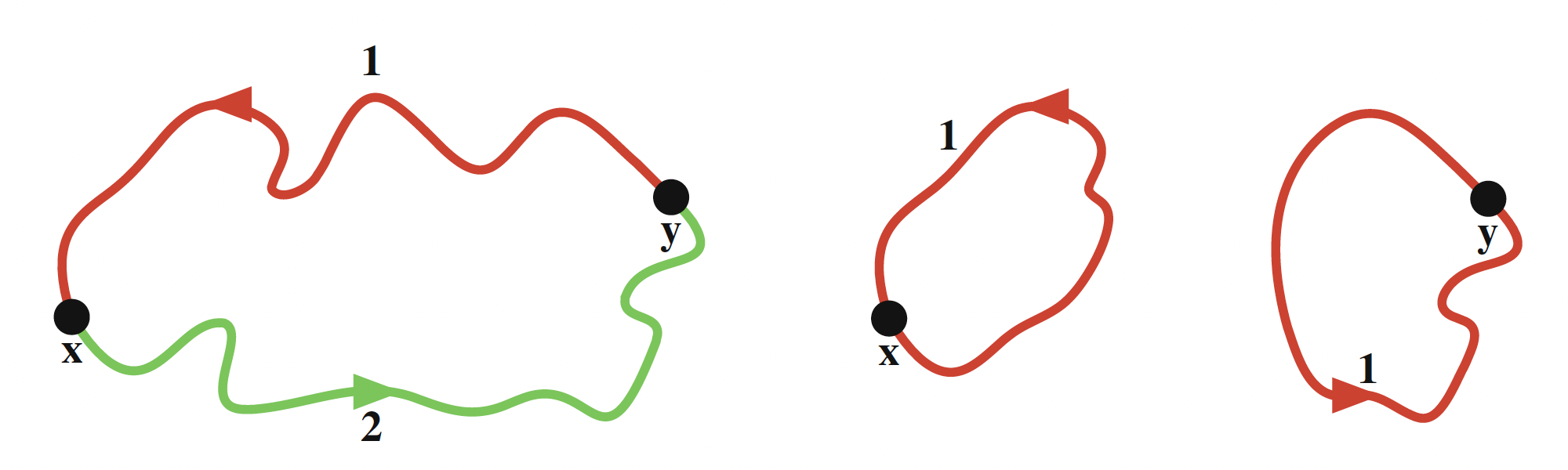}
    \caption{(\textit{left}) Illustration of a configuration that  contributes to $\langle Q^{12}_x Q^{21}_y \rangle$ \\
    (\textit{right}) a single colour index will provide lesser control over the range of configurations. Ref.\cite{thesis}}
    \label{G2l}
\end{figure}

The factor $n$ is included because a loop passing through $l$ and $l'$ can have any of the $n$ colours, whereas the correlator computes the probability for the process where this specific loop changes its colour from $1$ to $2$ and vice-versa. Similarly it follows that the choice for components $(1 \text{ and }2)$ in the above equation is arbitrary and any two distinct components will suffice.

Using the same idea as before, one can compute the probability for a single loop to pass through all the links $l_1, l_2, l_3, ... l_m$, in that order, as:
\begin{equation}
    G(l_1, l_2, l_3, ... l_m) = \frac{n}{A^m} \left \langle Q^{12}_{l_1} Q^{23}_{l_2}...Q^{m1}_{l_m} \right \rangle.
    \label{mpoint}
\end{equation}

\section{Joint length distribution \& its moments}

In the following, we discuss how the correspondence between correlations of $Q$ and geometrically defined events, discussed in the previous section, can be used to compute moments of the joint length distribution. Note however that the results to follow work only in the extended phase of the loop models. This is because the calculations consider that the dominant contribution in the correlations arises from the spatially constant part of $Q^{\alpha \beta} (x)$ i.e. the zero mode, which holds true only in the ordered phase when $\langle Q \rangle \neq 0$. The ordered phase of the $CP^{n-1}$ model corresponds to the extended phase of the loop models. [\cite{thesis}]

Label the length of loops $l_i$ in a configuration $C$, where $i=1,2...|C|$. Using the $m$-point correlation function Eq.($\ref{mpoint}$), one can show,
\begin{equation}
    \left \langle  \sum_i l^m_i   \right \rangle =
    \frac{n (m-1)!}{A^m} \int d^3x_1 d^3x_2..d^3x_m 
    \left \langle Q^{12}(x_1) Q^{23}(x_2)...Q^{m1}(x_m) \right \rangle.
    \label{momlen}
\end{equation}
In the lattice loop model setting, the integral over $d^3{x_k}$ implies one must sum over all links $x_k$ in the lattice. For a particular choice of links $(x_1,x_2,..x_m)$, the $m$-point correlation considers all the configurations $C$ which has a loop with that traverses the links $(x_1,x_2,..x_m)$ in that order. For such a given configuration $C$ and the loop $l_i$, we can evaluate the integrals by allowing $x_k$ to vary over the loop. The loop of length $l_i$ will contribute $l_i/(m-1)!$ where the combinatorial factor arises because the $m$ points $x_k$ follow a prescribed order. Eventually, the integral comes across every possible loop in every possible configuration, which is appropriately weighted by the correlator and we get $\left \langle  \sum_i l^m_i / (m-1)!  \right \rangle $. The factor $n/A^m$ carries over from Eq.(\ref{mpoint}).

In the ordered phase, the $m$-point correlation function is dominated by the zero mode of the field $Q^{\alpha \beta}(x) = Q_0 (z^{\alpha} \bar{z}^{\beta} - \delta^{\alpha \beta})$ and the trace operation is performed by averaging over all directions on the sphere $\mathbf{z}^{\dagger} \mathbf{z} = n$. The correlation function becomes
\begin{equation}
    \left \langle Q^{12}(x_1) Q^{23}(x_2)...Q^{m1}(x_m) \right \rangle = \text{Tr} \left ( |z^1|^2 |z^2|^2 ...|z^m|^2 \right )
    \label{momcalc}
\end{equation}
which is evaluated using the property
\begin{equation}
    \text{Tr} \left ( z^{\alpha_1} z^{\alpha_2}..z^{\alpha_q}
    \bar{z}^{\beta_1} \bar{z}^{\beta_2}...\bar{z}^{\beta_q}
    \right )
    = \frac{n^{q-1} n!}{(n+q-1)!} [z^{\alpha_1} z^{\alpha_2}..z^{\alpha_q}
    \bar{z}^{\beta_1} \bar{z}^{\beta_2}...\bar{z}^{\beta_q}]_G
    \label{A16}
\end{equation}
where $[..]_G$ indicates a Gaussian average that is computed by partitioning the product into all possible pairings using Wick's theorem and applying $[z^{\alpha} \bar{z}^{\beta}]_G = \delta^{\alpha \beta}$. The only non-vanishing pairing of components in Eq.(\ref{momcalc}) is given by
\begin{equation}
    \left [|z^1|^2 |z^2|^2 ...|z^m|^2\right ]_G = [z^1 \bar{z}^1]_G...[z^m \bar{z}^m]_G
\end{equation}
The correlation function thus becomes
\begin{equation}
        \left \langle Q^{12}(x_1) Q^{23}(x_2)...Q^{m1}(x_m) \right \rangle = Q_0^m \frac{n^{m-1} n!}{(n+m-1)!}.
\end{equation}
Using $\mathcal{L}_{\text{long}} = Q_o L^3 n /a$ and performing the integral in Eq.(\ref{momlen}) we get 
\begin{equation}
    \left \langle  \sum_i l^m_i  \right \rangle =
    \frac{ n! (m-1)!}{(n+m-1)!} \mathcal{L}_{\text{long}}^m
    \label{A19}
\end{equation}
In deriving the above equation, we considered $m$ to be an integer greater than one and the system to be in an ordered/extended loop phase. In this phase a finite fraction $f$ of the links lie on long loops which provide the dominant contribution in L.H.S. of Eq.(\ref{A19}) for $m>1$. The authors expect that the result can be analytically continued to $m=1$ when the sum is restricted to the long loops and we get $\sum_i^{'} l_i = \mathcal{L}_{\text{long}}$. This explains the choice of notation $\mathcal{L}_{\text{long}}$ in the R.H.S of the above equation which denotes the total length of all long loops. 

We can extend the ideas used to derive Eq.(\ref{momlen}) to calculate more general moments 
\begin{equation}
    C(m_1,m_2,...,m_q) = \mathcal{L}_{\text{long}}^{-\sum_k m_k} \left 
    \langle \sum^{'}_{i_1,..,i_q} l^{m_1}_{i_1} l^{m_2}_{i_2}... l^{m_q}_{i_q}      \right \rangle
\end{equation}
where the prime on the sum indicates that $i_1,...,i_q$ are distinct loops. Consider a ($\sum_k m_k$)-point correlation function 
\begin{equation}
    \left \langle
    \Gamma^{(1)} (x^{(1)}_1,x^{(1)}_2,...,x^{(1)}_{m_1}) \times... \times
    \Gamma^{(q)} (x^{(q)}_1,x^{(q)}_2,...,x^{(q)}_{m_q})
    \right \rangle
\end{equation}
which is a product of $q$ operators $\Gamma^{(k)}=$ each of which forces the points $x^{(k)}_1,...,x^{(k)}_{m_k}$ to lie on a single loop. For example, 
\begin{equation}
    \Gamma^{(1)} = Q^{12}(x^{(1)}_1) Q^{23}(x^{(1)}_2)...Q^{m_1 1}(x^{(1)}_{m_1})
\end{equation}
is simply the product of $m_1$ two-leg operators like in Eq.(\ref{momlen}). Following similar reasoning as before, the integral over the volume $ \Pi_{k,i_k} d^3 {x^{(k)}_{i_k}}$ will allow the points $x^{(k)}_{i_k}$ to vary on the loops and give 
\begin{equation}
     \frac{l^{m_1}_{i_1}}{ (m_1-1)!} \times ... \times \frac{l^{m_q}_{i_q}}{(m_q-1)!}.
\end{equation}
for each set of $q$ distinct loops $(i_1..i_q)$ in a configuration. The choice of $m_{\text{tot}} = \sum_k m_k$ points implies that the every valid configuration has $m_{\text{tot}}$ different color segments and thus the loops $l_{i_1},..,l_{i_q}$ necessarily have to be distinct. Once again employing Eq.(\ref{A16}) we get,
\begin{equation}
    \left \langle  \Gamma^{(1)} \times ... \Gamma^{(k)} \right \rangle  = Q_0^{m_{\text{tot}}} \frac{n^{m_{\text{tot}}-1} n!}{(n+m_{\text{tot}}-1)!}
\end{equation}
and the we find,
\begin{equation}
\left \langle \sum^{'}_{i_1,..,i_q} l^{m_1}_{i_1} l^{m_2}_{i_2}... l^{m_q}_{i_q}  \right \rangle =  \mathcal{L}_{\text{long}}^{m_{\text{tot}}}\frac{n^q \Gamma(n) \Gamma(m_1) \Gamma(m_2) ..\Gamma(m_q) }{\Gamma(n+m_{\text{tot}})}
\end{equation}
as the general moment $C(m_1,m_2,...,m_q)$ for the joint length distribution.


\begin{thebibliography}{}
 
 \bibitem{resumm}
 Nisheeta Desai and Sumiran Pujari, "Resummation-based updates for Stochastic Series Expansion Quantum Monte Carlo", 2021. arXiv: 2104.06792v1
 
 \bibitem{ueltarxiv}
 Daniel Ueltschi, "Quantum Heisenberg models and random loop representations", arXiv:1211.4141
 
 \bibitem{ueltjmp}
 Daniel Ueltschi, "Random loop representations for quantum spin systems", J. Math. Phys. 54, 083301 (2013)
 
 \bibitem{sandvik1}
 Anders W. Sandvik, "Computational Studies of Quantum Spin Systems", arXiv:1101.3281
 
  \bibitem{sandvik2}
 Anders W. Sandvik, "Stochastic Series Expansion Methods", arXiv:1909.10591
 
 \bibitem{thesis}
 Adam Nahum, "Critical Phenomena in Loop Models", Springer Theses Series.
 
 \bibitem{prl1}
 Adam Nahum, J.T. Chalker \textit{et al}. "Length Distributions in Loop Soups", Phys. Rev. Lett., 111, 100601 (2013)
 
 \bibitem{prl2}
 Adam Nahum, J.T. Chalker, P. Serna, M. Ortuno and A. M. Somoza, "3D Loop Models and the $CP^{n-1}$ Sigma Models", Phys. Rev. Lett. 107, 110601, 
 \bibitem{prb}
 Adam Nahum, J.T. Chalker, P. Serna, M. Ortuno and A. M. Somoza, "Phase transitions in three-dimensional loop models and the $CP^{n-1}$ sigma model", Phys. Rev. B 88, 134411
 
 \bibitem{prx} 
 Adam Nahum, J.T. Chalker, P. Serna, M. Ortuno and A. M. Somoza, "Deconfined Quantum Criticality, Scaling Violations, and Classical Loop Models", Phys. Rev. X 5, 041048
 
 \bibitem{data}
 Peter Young, "Everything you wanted to know about Data Analysis and Fitting but were afraid to ask", (2012) arXiv:1210.3781

 \end{thebibliography}

\thispagestyle{plain}

\begin{center}
 \Large {\bf \uppercase{Acknowledgements}}
\end{center}

\vspace{3\baselineskip}

\noindent
I would like to extend my deepest gratitude to my supervisor Prof. Sumiran Pujari for his constant guidance and support throughout the course of this project. I am also very grateful to Dr. Nisheeta Desai for her aid through numerous valuable discussions. 

\noindent
\vspace{\baselineskip} \\
\textbf{\myname} \\
IIT Bombay \\
$19^{\text{th}}$ August 2022

\end{document}